\documentclass[aps,prd,floats,nofootinbib]{revtex4}

\usepackage{latexsym}
\usepackage{amssymb}
\usepackage[dvips]{graphicx}
\usepackage[applemac]{inputenc}
\usepackage[nottoc]{tocbibind} %prevents the entry "Table of Contents" in the "Table of Contents"
\usepackage{float}
\usepackage{fancyhdr}
\usepackage{amsfonts}
\usepackage{amsmath}
\usepackage{color}
\usepackage{mathrsfs}
\usepackage{bm}% bold math
\usepackage{epsfig}% bold math
\usepackage{feynmf}

\newcommand{\be}{\begin{equation}}
\newcommand{\ee}{\end{equation}}

\def\lsim{\mathrel{\raise.3ex\hbox{$<$\kern-.75em\lower1ex\hbox{$\sim$}}}} 
\def\gsim{\mathrel{\raise.3ex\hbox{$>$\kern-.75em\lower1ex\hbox{$\sim$}}}}

\def\beq{\begin{eqnarray}}
\def\eeq{\end{eqnarray}}
\def\bea{\begin{eqnarray}}
\def\eea{\end{eqnarray}}

\begin{document}

%\title{Light WIMP Dark Matter in Light of CoGeNT \\or\\ What If CoGeNT, DAMA and CDMS Are Seeing Dark Matter?}

\title{Implications of CoGeNT and DAMA for Light WIMP Dark Matter}
\author{A. Liam Fitzpatrick$^{a}$, Dan Hooper$^{b,c}$ and Kathryn M. Zurek$^{d}$}

\address{
${^a}$ Department of Physics, Boston University, Boston, MA, USA\\ $^b$ Particle Astrophysics, Fermi National Accelerator Laboratory, Batavia, IL 60563, USA \\ $^c$ Department of Astronomy and Astrophysics, University of Chicago, Chicago, IL 60637, USA \\ ${^d}$ Michigan Center for Theoretical Physics and Department of Physics,  University of Michigan, Ann Arbor, Michigan 48109, USA}

\begin{abstract}

In this paper, we study the recent excess of low energy events observed by the CoGeNT collaboration, and discuss the possibility that these events originate from the elastic scattering of a light ($m_{\rm DM }\sim$5-10 GeV) Weakly Interacting Massive Particle (WIMP). We find that such a dark matter candidate may also be capable of generating the annual modulation reported by DAMA, without conflicting with the null results from other experiments, such as XENON10.  The regions implied by CoGeNT and DAMA are also near those required to produce the two observed CDMS events.  A dark matter interpretation of the CoGeNT and DAMA observations favors a region of parameter space that is especially attractive within the context of Asymmetric Dark Matter models. In such models, the cosmological dark matter density arises from the baryon asymmetry of the universe, naturally leading to the expectation that $m_{\rm DM }\sim$1-10 GeV. We also discuss neutralino dark matter from extended supersymmetric frameworks, such as the NMSSM. Lastly, we explore the implications of such a dark matter candidate for indirect searches, and find very encouraging prospects for experiments attempting to detect neutrino or gamma ray annihilation products.

%We study the implications of the recent excess of low energy events observed by CoGeNT for light WIMP dark matter, including compatibility of the CoGeNT window with null results from XENON10 and with the DAMA annual modulation signal.  We find that CoGeNT is consistent with the XENON10 null result at 90\% C.L. using new measurements of the quenching factor, and show that with an appropriate choice of halo model it is also consistent with the DAMA signal in the low mass elastic scattering window.  We discuss how the CoGeNT window favors a region predicted by asymmetric dark matter models, where the dark matter density arises from the baryon asymmetry, and also consider dark matter in the CoGeNT window arising in the NMSSM.  We then turn to carrying out an operator analysis of the light WIMP window, and discuss implications for indirect detection.  

\end{abstract}

\maketitle

%%%%%%%%%%%%%%%%%%%%%%%%%%%%%%%%%%%%%%%%%%%%%%%%%%%%%%%%%%%%%%%%%%%%%%

\section{Introduction}

The direct and indirect detection of dark matter has recently become a topic of intense discussion and activity, with a number of reported signals having been interpreted as possible detections of particle dark matter. Among direct detection experiments, the DAMA collaboration, which has collected over 1.17 ton-years of data, reports an annual modulation in their event rate with 8.9$\sigma$ significance, which they interpret as evidence of dark matter~\cite{DAMAnew}. The CDMS experiment has also observed a small excess (2 events, corresponding to less than $2 \sigma$ significance) in their low energy nuclear recoil window~\cite{Ahmed:2009zw}. Furthermore, the CoGeNT collaboration has very recently announced the observation of an excess of events at low energies~\cite{cogentnew}. If the CoGeNT excess is interpreted as a detection of dark matter, this points to dark matter with a mass in the range of approximately $m_{\rm DM}\sim$5-10 GeV, and an elastic scattering cross section with nucleons of approximately $\sigma \sim 7 \times 10^{-41}$ cm$^2$. Remarkably, a dark matter candidate with this approximate mass and cross section would also be capable of producing both the two events observed by CDMS, and the modulation signal reported by DAMA.
% (when proposed effects of channeling are taken into account).
% region, when a 
%%xxx-liam
%%channeling effect is taken into account.
%proposed channeling effect is included.
%%xxx

Indirect detection efforts have also recently produced a number of tantalizing, if unconfirmed, signals of dark matter.  The PAMELA experiment has reported an excess of cosmic ray positrons between 10 and 100 GeV relative to the predictions of standard cosmic ray models~\cite{PAMELA}. The Fermi Gamma Ray Space Telescope (FGST) collaboration has also reported a significantly harder spectrum of electrons (plus positrons) than had been anticipated~\cite{Fermi}.  In addition, there are apparent excesses in radio and gamma rays originating from the inner kiloparsecs of the Milky Way: the WMAP haze~\cite{haze}, and the recently reported Fermi (gamma ray) haze~\cite{fermihaze}.  Lastly, it has recently been shown that the spectrum and angular distribution of gamma rays from the Galactic Center, as observed by the FGST, is consistent with an annihilating dark matter interpretation~\cite{lisa}. Although each of these observations is consistent with annihilating dark matter, it is not yet possible to rule out less exotic astrophysical origins, such as pulsars~\cite{pulsars}.

%Given the apparent discovery of DM through direct detection, and the high sensitivity of gamma ray experiments such as Fermi to DM annihilation, there are many avenues for exploring its properties.  

In this paper, we focus on recent results from direct detection experiments, and examine the implications of the detections made by CoGeNT, DAMA, and CDMS. \footnote{The dark matter candidates giving rise to direct detection signals considered here could not generate the positron excesses in Fermi and PAMELA, though multiple components of dark matter could simultaneously give rise to multiple signals of different type in direct and indirect detection experiments \cite{multicomp}.} We begin by discussing the compatibility of these observations with each other, and with the null results of other experiments; in particular the recent low threshold analysis of the XENON10 collaboration~\cite{XENONinelastic}. We then turn to a discussion of dark matter models that are naturally able to accommodate the observations of CoGeNT, DAMA, and CDMS.  Particularly attractive within this context are Asymmetric Dark Matter (ADM) models \cite{ADMmodels,ADM}, in which the dark matter abundance is related to the baryon asymmetry of the universe. In such models, the dark matter's mass is predicted to be a factor of several times heavier than the proton; precisely within the region indicted by CoGeNT. We also explore supersymmetric models in which these signals can be produced. In particular, we discuss neutralinos within the context of the next-to-minimal supersymmetric standard model (NMSSM).  Lastly, we turn our attention to the prospects for indirect dark matter searches. We find that, while the region implied by CoGeNT is not ruled out by any indirect detection experiments, the outlook for dark matter's indirect detection is very encouraging in this scenario. In particular, neutrino and gamma ray searches for dark matter annihilation products are currently within a factor of $\sim$$2$ of the sensitivity required to test many of the models capable of generating the signals reported by CoGeNT, DAMA and CDMS.

% for direct and indirect detection.  We begin by discussing the compatibility of CoGeNT with the results of other null experiments, in particular the recent low threshold analysis of XENON \cite{XENONinelastic}, and the annual modulation observed by DAMA \cite{DAMAnew}.  At face value, the detections of CoGeNT and DAMA seem in contradiction with this constraint, and we discuss 
%%xxx-liam
%% how the tension can be alleviated.
%what assumptions must be relaxed for the tension to be alleviated.
%%xxx

%\begin{figure}[htp]
%\label{fig:decaychain}
%\centering
%\includegraphics[width=0.5\textwidth]{FigSchematicNeq1.pdf}
%\caption{ Schematic representation of the $n=1$ class of processes considered in this work, with additional USR.  The parent particle is the state which decays to the visible SM particles and the child DM particles. }
%\end{figure}

\section{CoGeNT, DAMA, CDMS and XENON}
%with Null Results and the DAMA Modulation Signal}

In this section, we explore whether a dark matter interpretation of the CoGeNT excess is consistent with the signals reported by DAMA and CDMS, and whether such an interpretation is consistent with the null results of XENON10~\cite{xenonnull} and the CDMS silicon analysis~\cite{CDMSIISi,CDMSIISi2}. In particular, we follow Refs.~\cite{petriello,Chang,Fairbairn,savage} in studying direct detection signals in the low mass WIMP region, including the possible effects of channeling in the DAMA experiment~\cite{channeling}.

%We begin by exploring the compatibility of CoGeNT with the null results of the XENON and CDMS experiments, and discuss compatibility of DAMA with the low mass WIMP region \cite{petriello,savage} opened by channeling \cite{channeling}.%

\subsection{Detection Rates}

%%%xxx-liam - We're not considering inelastic.
%For generality, we consider both elastic and inelastic scattering of DM off nuclei.  We will find that the largest region of parameter space is available for elastic scattering dark matter.
%xxx
The differential rate of dark matter elastic scattering events per unit detector mass in nuclear recoil energy is given by
\be
\frac{dR}{dE_R} = N_T \frac{\rho_{DM}}{m_{DM}} \int_{|\vec{v}|>v_{\rm min}} d^3v\, vf(\vec{v},\vec{v_e}) \frac{d\sigma}{d E_R},
\label{rate1}
\ee
where $N_T$ is the number of target nuclei per unit mass, $m_{DM}$ is the dark matter particle mass, $\rho_{DM}= 0.3\, {\rm GeV/cm^3}$ is the local dark matter density, $\vec{v}$ is the dark matter velocity in the frame of the Earth, $\vec{v_e}$ is the velocity of the Earth with respect 
to the galactic halo, and $f(\vec{v},\vec{v_e})$ is the distribution function of dark matter particle velocities, which we take to be 
a standard Maxwell-Boltzmann distribution:
\be
f(\vec{v},\vec{v_e}) = \frac{1}{(\pi v_0^2)^{3/2}} {\rm e}^{-(\vec{v}+\vec{v_e})^2/v_0^2}.
\ee
The Earth's speed relative to the galactic halo is $v_e=v_{\odot}+v_{\rm orb}{\rm cos}\,\gamma\, {\rm cos}[\omega(t-t_0)]$ with $v_{\odot}=v_0+12\,{\rm km/s}$, 
$v_{\rm orb}=30 {\rm km/s}$, ${\rm cos}\,\gamma=0.51$, $t_0={\rm June \, 2nd}$, and $\omega=2\pi/{\rm year}$.  We set 
%xxx-liam
%$v_0={\rm 230 \,km/s}$
$v_0={\rm 270 \,km/s}$
%xxx
 throughout most of our analysis.  The upper limit 
of the velocity integration of  Eq.~(\ref{rate1}) is the galactic escape velocity, ${\rm 490\,km/s} \leq v_{\rm esc} \leq {\rm 730\,km/s}$ at 90\% C.L.~\cite{Kochanek:1995xv}, and  
the minimum dark matter 
velocity, $v_{\rm min}$, is
\be
v_{\rm min} = \sqrt{\frac{E_R m_N}{2 \mu^2_1}}.
\ee
where $E_R$ is the recoil energy, $m_N$ is the mass of the target nucleus,
and $\mu_1$ is the dark matter-nucleus reduced mass.
For a spin-independent cross section between dark matter particles and nuclei, we have~\cite{Jungman:1995df} 
\be
\frac{d\sigma}{d E_R} = \frac{m_N}{2 v^2} \frac{\sigma_N}{\mu_n^2} \frac{\left[f_p Z+f_n (A-Z)\right]^2}{f_n^2} F^2(E_R),
\label{cross1}
\ee
where $\mu_n$ is the reduced mass of the dark matter particle and nucleon (proton or neutron), $\sigma_N$
 is the scattering cross section of the dark matter 
particle with 
neutrons, $Z$ and $A$ are the proton and atomic numbers of the nucleus, and $f_{n,p}$ are the coupling strengths of the dark matter particle to neutrons and protons respectively.  The couplings $f_{n,p}$ are 
calculated from a coherent sum over the couplings to the quark constituents of the nucleon. We use the Woods-Saxon form factor, $F(E_R)$~\cite{Gondolo,Fricke}.

\begin{figure}[t]
\centering
\includegraphics[width=3.5in]{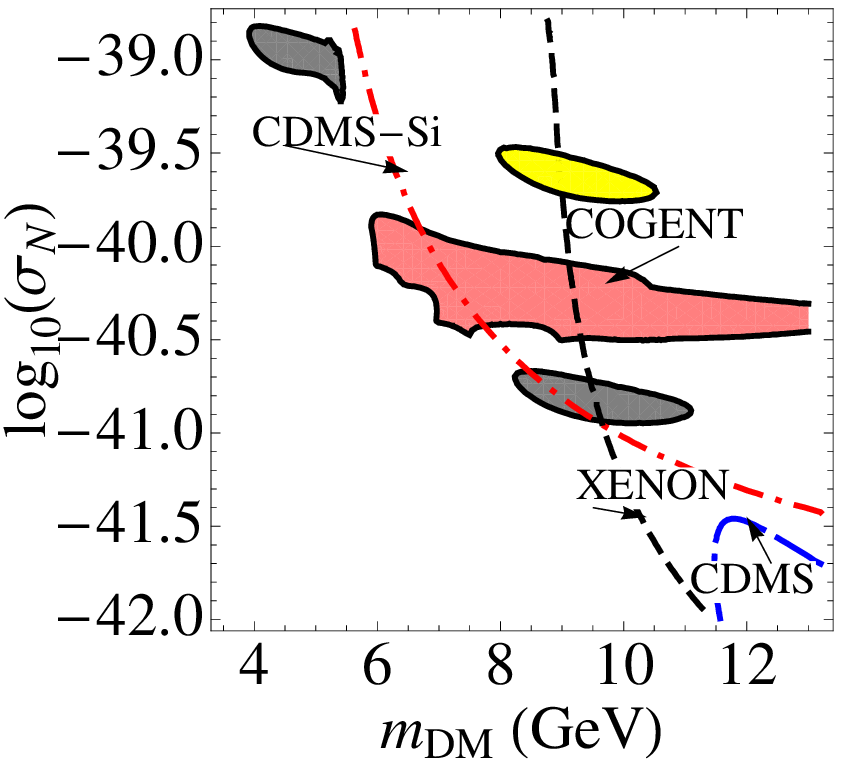}
\includegraphics[width=3.5in]{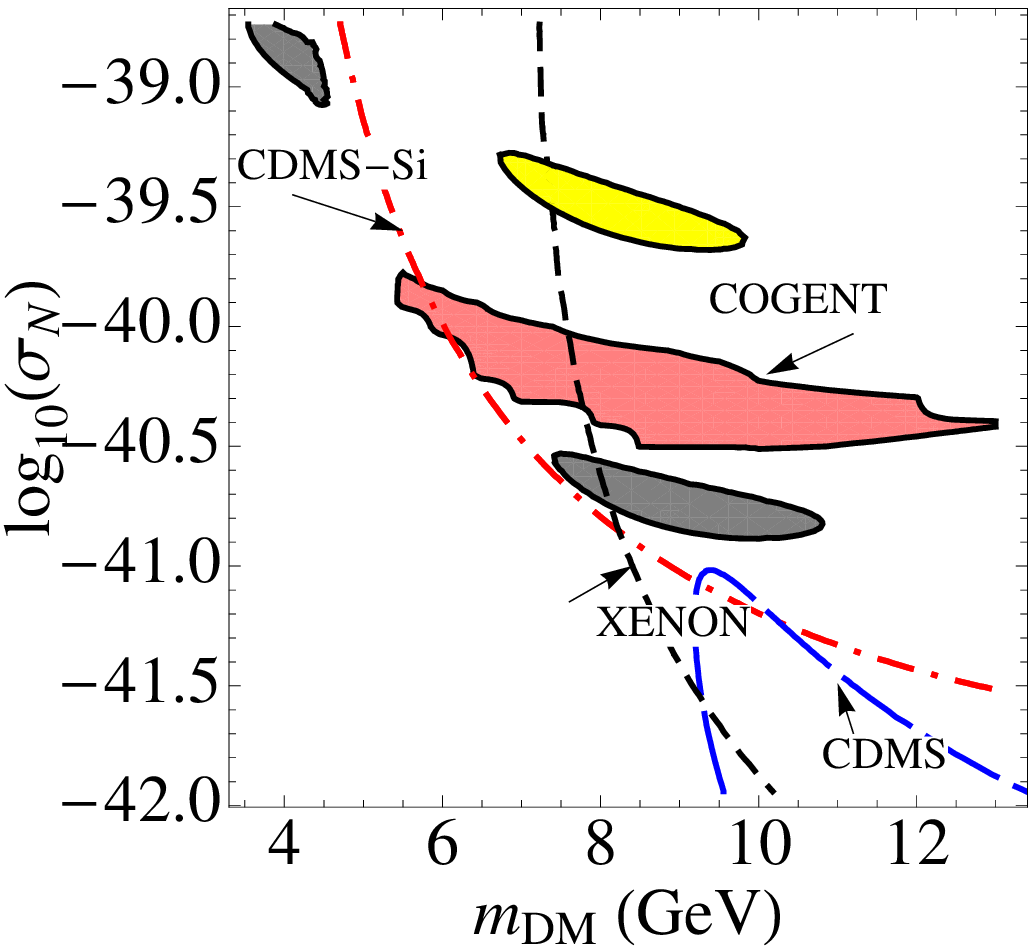}
\caption{The regions in the elastic scattering cross section (per nucleon), mass plane in which dark matter provides a good fit to the CoGeNT excess, compared to the region that can generate the annual modulation reported by DAMA at 90\%
confidence (darker grey regions). In this figure, we have adopted $v_0=270$ km/s and use two values of the galactic escape velocity: $v_{\rm esc}=490$ km/s (left) and $v_{\rm esc}=650$ km/s (right). In calculating the DAMA region, we have 
neglected the lowest energy bin (the effect of this is shown in later figures)
and treated channeling as described in Ref.~\cite{Bernabei:2007hw}. If a smaller fraction of events are channeled in DAMA than is estimated in Ref.~\cite{Bernabei:2007hw}, the DAMA region will move upward, toward the yellow regions (near $\sigma_N\approx 10^{-39.5}$ cm$^2$, which include no effects of channeling), improving its agreement with CoGeNT. Also shown is the 90\% C.L. region in which the 2 events observed by CDMS can be produced. If the escape velocity of the galaxy is taken to be relatively large, this region can also approach those implied by CoGeNT and DAMA. Constraints from the null results of XENON10 and the CDMS silicon analysis are also shown. For the XENON10 constraint, we have used the lower estimate of the scintillation efficiency (at $1\sigma$) as described in Ref.~\cite{Leff}.}
\label{fig1}
\end{figure}

CoGeNT is an ultra-low noise detector operating in the Soudan mine, consisting of 0.33 kg (fiducial mass) of germanium. Over a period of 56 days, the CoGeNT detector recorded approximately $\sim$100 events (with characteristics corresponding to interactions in the bulk of the crystal) above their expected background with ionization energy between 3.2 keVee\footnote{Ionization energies are given in units of keVee, or keV electron equivalent. For a 1 keV nuclear recoil, for example, the equivalent electron energy (in keVee) is $q_x \times 1 \,{\rm keV}$, where $q_x$ is the quenching factor for the nuclear material composing the scintillator. For the germanium target used in CoGeNT, the quenching factor is approximately 0.2 to 0.3 for events in the energy range of the observed excess. For the materials composing the DAMA detectors, $q_{Na} \approx 0.3$ and $q_I \approx 0.09$.} and their threshold of 0.4 keVee~\cite{cogentnew}. The CoGeNT collaboration has pointed out that their events can be fit very well by a $\sim$10 GeV dark matter particle with an elastic scattering cross section of $\sim 7 \times 10^{41}$ cm$^2$. In Fig.~\ref{fig1}, we confirm this conclusion, where we show the parameter space region in which elastically scattering dark matter can accommodate the CoGeNT excess
%xxx-liam
at 90\% confidence.
%xxx
In this figure, we have used $v_0=270$ km/s and $v_{\rm esc}=490$ km/s.  
%xxx-liam
Here, and throughout this paper, 90\% (99\%) confidence regions are defined
as contours of $\chi^2=\chi^2_{\rm min} + 4.61 \ (9.21)$, while
constraints from null experiments are defined as 90\% limits based on
the maximum gap~\cite{Yellin} method.  To carry out this fit, we have assumed that the background is well described by an exponential plus constant, and we have required bin-by-bin that the background not exceed the amplitude of the dark matter signal.
% (in Figs~2 and 3 for illustration of the effect of the amplitude of the background, we allow the background component to be as large as the signal bin-by-bin).  
Without a constraint on dark matter signal to background, the entire spectrum is well fit by a pure exponential background.  Tighter constraints on the amplitude of the background will correspond to the dark matter signal region shifting to larger cross sections.  We fit the data in 0.05 keV-electron-equivalent (keVee) bins from threshold at 0.4 keVee to  1.8 keVee where the dark matter signal is negligible.  Peaks in the data (consistent with a background from radioactive tin) at 1.1 and 1.29 keVee are fit by Gaussians of relative height $0.4$ and with width  consistent with the experimental resolution at those energies (0.0774 and 0.078 keVee respectively).
%xxx
We can see that for appropriate choices of the halo model and the fraction of channeled events in DAMA, the CoGeNT region can be consistent at 90\%~C.L. with the DAMA signal and the null results XENON and CDMS-Si.  Some consistency between the preferred region for CDMS with DAMA and CoGeNT can also be found.  We now turn to discussing in detail how this occurs.

The DAMA experiment \cite{DAMAnew} observes an annual modulation in their count rate, which can be parameterized as 
\be
R_i = R_i^0 + S_i^1 {\rm \cos} [\omega(t - t_0)].
\ee
%xxx-liam
%where $t_0={\rm June \, 2nd}= 152.5 \mbox{ days}$ is the time when the Earth is moving with its maximum speed with respect to the galactic halo and $T = 2 \pi/\omega = 1 \,{ yr}$. 
%xxx
   The subscript $i$ in this expression denotes different energy bins.   The constant term $R_i^0$ is composed of both a signal component coming from dark matter initiated processes, and a background component arising from other sources of nuclear recoil: $R_i^0 = b_i^0+S_i^0$.    The expressions for $S_i^0$ and $S_i^1$ are obtained by integrating Eq.~(\ref{cross1}) over a given energy bin. 

%xxx-liam
%An important effect which must be accounted for when interpreting DAMA's signal, and especially when comparing this signal to the results of other direct detection experiments, is channeling.
Channeling is a potentially important but difficult-to-predict theoretical effect which can significantly change the interpretation of DAMA's signal, especially 
when comparing this signal to the results of other direct detection experiments.
%xxx
 In a typical nuclear recoil event, only a fraction of the total energy is detected (as a combination of scintillation light, heat, and ionization, depending on the detector). The ratio of the observed energy to the total recoil energy is known as the quenching factor. For crystal scintillators, however, such as those used by the DAMA collaboration, a portion of the events will be ``channeled", causing 
%xxx-liam
%essentially all 
most
%xxx
of the recoil energy in those events to be observed (effectively changing the quenching factor to $q \approx 1$). This occurs when the incident particle interacts only electromagnetically with the scintillator material, which can occur for certain energies and incidence angles. The importance of this effect for the DAMA experiment was first discussed in Ref.~\cite{channeling}, and an analysis of its effect was performed by the DAMA collaboration in Ref.~\cite{Bernabei:2007hw}; we refer the reader to these references for a more detailed discussion. 

Questions of how to quantitatively account for the effects of channeling in DAMA have made efforts to interpret their results, and to compare their results to those of other experiments, somewhat difficult. In particular, the compatibility of the DAMA modulation with the CoGeNT and CDMS excesses (and the compatibility of DAMA with the null results of XENON and other searches) depends strongly on the fraction of elastic scattering events which are channeled rather than quenched. An estimate of the fraction of channeled events based on 
simulations is given in Fig.~4 of Ref.~\cite{Bernabei:2007hw}.  We use the following simple parameterization  in our analysis~\cite{Foot:2008nw}:
\be
f_{Na} \approx \frac{1}{1+1.14 E_R(\mbox{keV})} ,\,\,\,\,\,\,\,\,\,\, \mbox{        } f_{I} \approx \frac{1}{1+0.75 E_R(\mbox{keV})} .
\label{channeldfraction}
\ee
%xxx-liam
It should be noted, however, that channeling at DAMA has not been experimentally
verified, and studies of NaI at energies above the region of interest for
DAMA have failed to find such an effect~\cite{Graichen}.
%xxx-liam
  We also consider
the DAMA preferred region if channeling is not present.  In this case,
scattering off of iodine cannot reproduce the DAMA signal at the low masses
of interest.
Scattering off of sodium is relevant, however, and is shown as a yellow region in each frame of Fig.~\ref{fig1}.
The cross sections required to explain DAMA without any channeling are somewhat too high to explain CoGeNT,
but the range of masses needed to explain the two experiments is surprisingly
similar.  
%xxx
%xxx

In Fig.~\ref{fig1}, we show along side the CoGeNT region the parameter space that can accommodate DAMA's annual modulation signal.  In addition to the DAMA and CoGeNT regions, we also show the region of the parameter space in which the 2 events observed by CDMS can be generated (consistent with the small excess recently reported~\cite{Ahmed:2009zw}).
%xxx-liam
%In Fig.~1(a) we see how modifying the channeled fraction to 60\% of its estimated value in Eq.~\ref{channeldfraction} easily moves the DAMA region into better agreement with CoGeNT.
 If the actual fraction of channeled events is smaller than its estimated
value in Eq.~\ref{channeldfraction}, then the DAMA region simply
shifts upward in cross section to accommodate the change.
Thus, in Fig.~1(a), one can see by eye how modifying the channeled fraction 
to 60\% of its estimated value easily moves the 
DAMA region into better agreement with CoGeNT.
%xxx
   In Fig.~1(b) we also see how $v_0$ shifts the DAMA region to lower masses.
%xxx-liam
% We next discuss how changes in the scintillation efficiency, $v_0$ and $v_{esc}$
The two constraint plots in Fig.~1 are considerably more favorable
for a light elastically scattering WIMP interpretation of DAMA
than previous results~\cite{Kopp:2009qt}. The reason for this being that the allowed region of parameter space depends quite sensitively
on assumptions about the halo model and the XENON10 scintillation efficiency.
We next turn our attention to these details and how they impact the constraints from CDMS and XENON10.

%We wish to study the consistency of the CoGeNT signal with the previously reported signals from DAMA (and the two events observed in CDMS).  We show in Table~(\ref{ExpFacts}), the characteristics of the relevant experiments for the low mass window consistent with the CoGeNT, DAMA and CDMS windows.  We now turn to a detailed analysis of the allowed parameter space and the consistency with null experiments.  The constraints from XENON will be the most important for alleviating tension between CoGeNT and DAMA with the null experiments.%

\subsection{Consistency with Null Results}

In the light WIMP window, the null results of XENON10 are among the most constraining, due to its relatively low threshold of 4.5 keV.  Recently that threshold was lowered further to 2 keV~\cite{XENONinelastic}, although no constraint was derived on low mass, elastically scattering dark matter in that analysis. To derive the XENON10 constraint, software efficiencies given in Ref.~\cite{XENONinelastic} are coupled with an $S_2$ twelve electron efficiency and an $S_1$ detector acceptance efficiency taken from Ref.~\cite{sorensen}. At face value, the XENON10 constraint appears to completely rule out the regions of the low mass window favored by DAMA and CoGeNT. 

\begin{figure}[htp]
\centering
\includegraphics[width=3.5in]{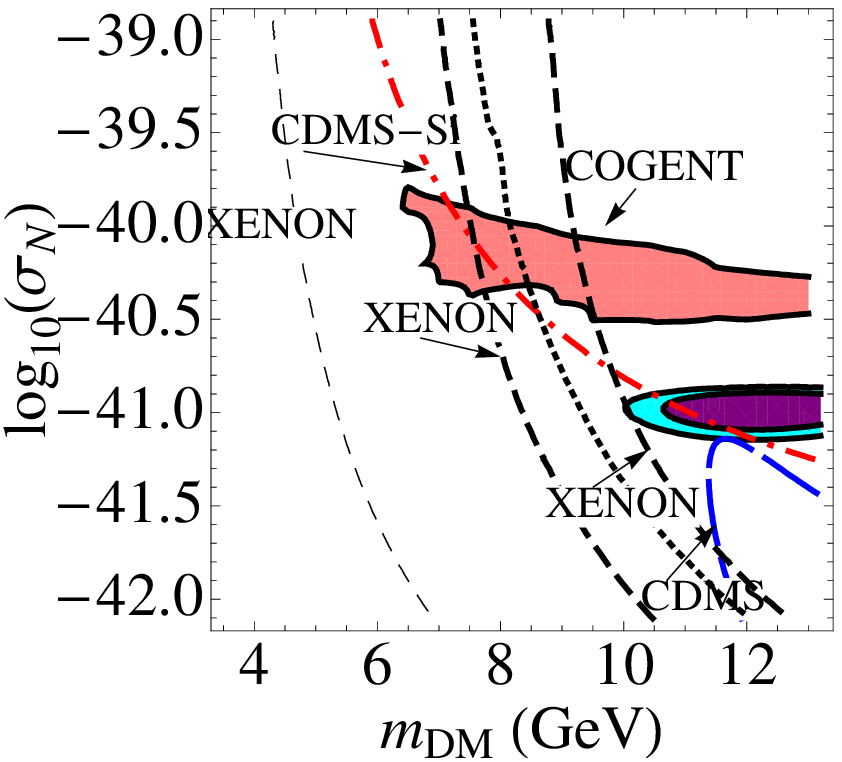}
\includegraphics[width=3.5in]{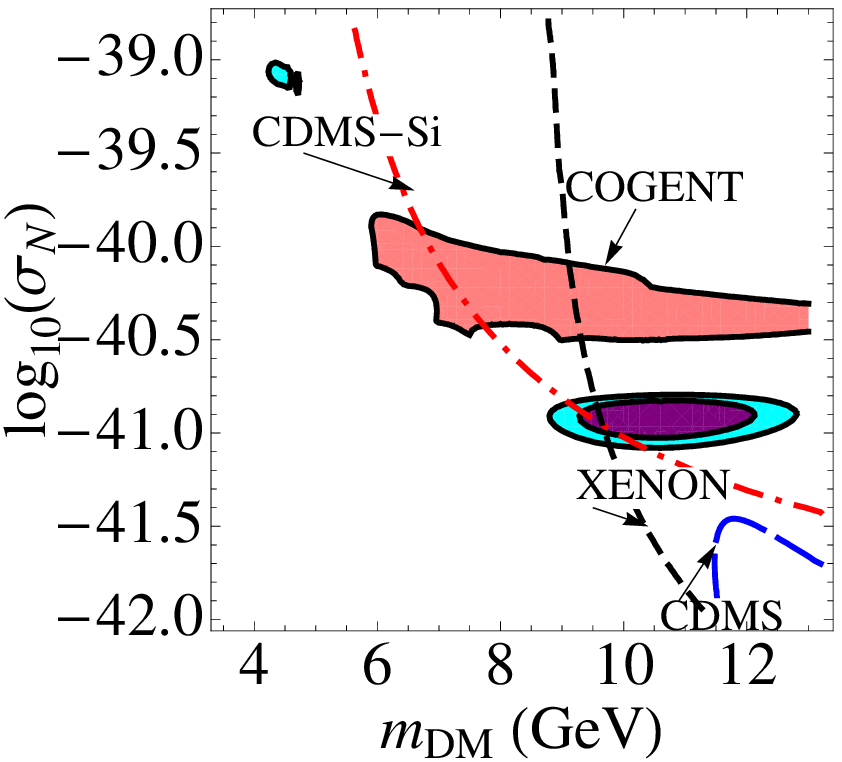}
\caption{{\em Left panel:} To illustrate the effect of uncertainties in the
XENON10 scintillation factor, $L_{\rm eff}$, we show the CoGeNT and DAMA allowed regions alongside constraints from the new XENON10 analysis using $L_{\rm eff}=0.19$ (light gray dashed), and $L_{eff}$ at the central and lower $1\sigma$ values from the new measurement \cite{Leff} (the two thick black dashed curves). 
In between the central and lower $1\sigma$ $L_{eff}$ curves
we have taken $-\frac{1}{2} \sigma$ $L_{eff}$ values in only the three energy
bins near XENON10's threshold (black dotted).   We also show constraints from the CDMS silicon analysis (red dot-dashed), and the region in which the two events observed by CDMS can be produced (blue long dashed). Here, we have used $v_0=220$km$/$s and $v_{\rm esc}=500$km$/$s. With the lower values of $L_{\rm eff}$ values, the tension between XENON10 and CoGeNT is alleviated. {\em Right panel:}  
To illustrate the effect from changing the halo model, CoGeNT, CDMS, XENON10, and DAMA results are shown, but with $v_0 = 270$km$/$s and $v_{\rm esc}$=490km$/$s.}
\label{fig:noLeff}
\end{figure}

There are reasons to be skeptical of such a conclusion, however. In particular, a new measurement was recently made of the scintillation efficiency (the fraction of nuclear recoil energy that goes into scintillation light) of liquid xenon, $L_{\rm eff}$~\cite{Leff}.  According to the new measurement, the scintillation efficiency is significantly lower than previously reported by the XENON10 collaboration, who take $L_{\rm eff} = 0.19$ in their analysis. A lower scintillation efficiency translates to a reduced sensitivity of xenon-based detectors to light dark matter. 
%In the right frame of Fig.~\ref{fig:noLeff}, we show the impact of this measurement on the region excluded by XENON10. From the figure, we see that the new measurement shifts the region excluded by XENON10 to significantly higher masses. Despite this shift,
Although the central values of the new $L_{\rm eff}$ measurement still lead to considerable tension with the region implied by CoGeNT, if we adopt values of $L_{\rm eff}$ which are $1 \sigma$ below its central values, we find that consistent regions can be found. This constraint is included in Fig.~\ref{fig:noLeff}, along with the constraint from the CDMS silicon analysis~\cite{CDMSIISi,CDMSIISi2}.
We show in Fig.~\ref{fig:noLeff}, for comparison, the constraints from XENON10
using the old $L_{\rm eff}=0.19$, along side those using the new measurement
with and without taking into account the $1 \sigma$ uncertainty in each
bin.  Note that the XENON10 excluded region changes rather drastically
with $L_{\rm eff}$.  Although this allows us to conclude that the CoGeNT region is consistent with the XENON10 constraint, the region favored by DAMA still appears to be favored by XENON10. The most important measurements of $L_{\rm eff}$ 
are in the
energy bins near XENON10's threshold, since these determine the lowest
recoil energies that XENON10 can probe.  For illustration, we also show
in Fig.~\ref{fig:noLeff} the constraint 
taking the $-\frac{1}{2} L_{\rm eff}$ values only in
the three bins centered on 4.9, 5.7, and 6.4 keV (and $-\frac{1}{4}L_{\rm eff}$
at 3.9 keV so that scintillation always increases with recoil energy), and
central values elsewhere.

%inconsistent with XENON10 at the 99\% C.L.  

The second significant effect that we have taken into account is shifts
from uncertainty in the halo model.  It is generally considered that
$v_0=220$km$/$s is a useful starting point for the halo model; however, if the 
distribution of velocities is broader, then lighter dark matter will be 
favored, bringing the DAMA region below the XENON10 constraint curve.
Taking $v_0 = 270 \mbox{ km/s}$, we see in the right frame of Fig.~\ref{fig:noLeff} that the region consistent with DAMA shifts to lower masses, and becomes consistent with both the constraints of the null experiments and with the CoGeNT preferred region.

\begin{figure}[htp]
\centering
\includegraphics[width=3.5in]{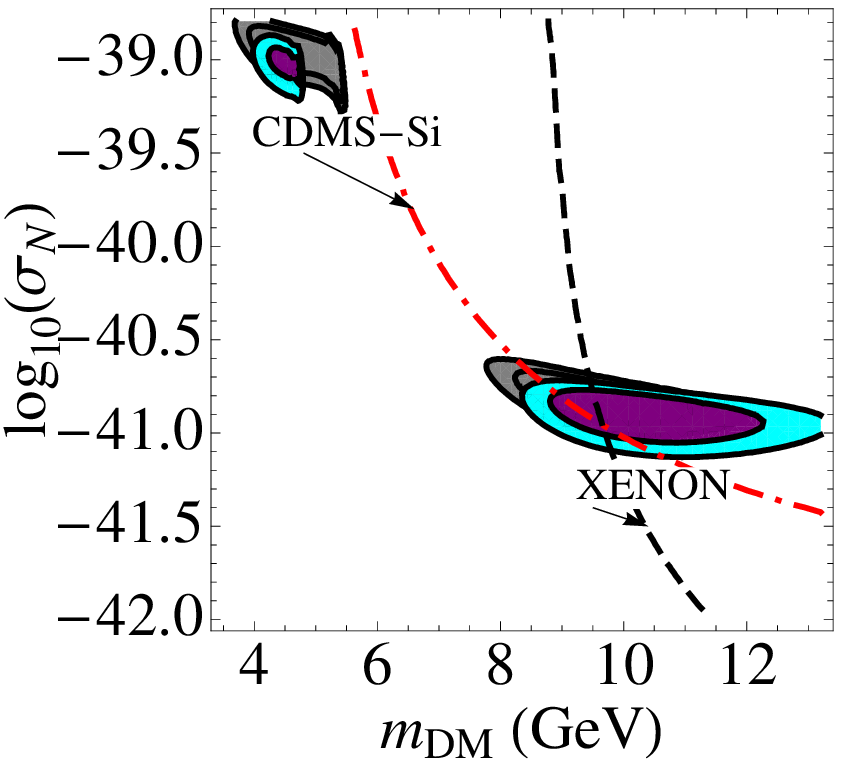}
\includegraphics[width=3.5in]{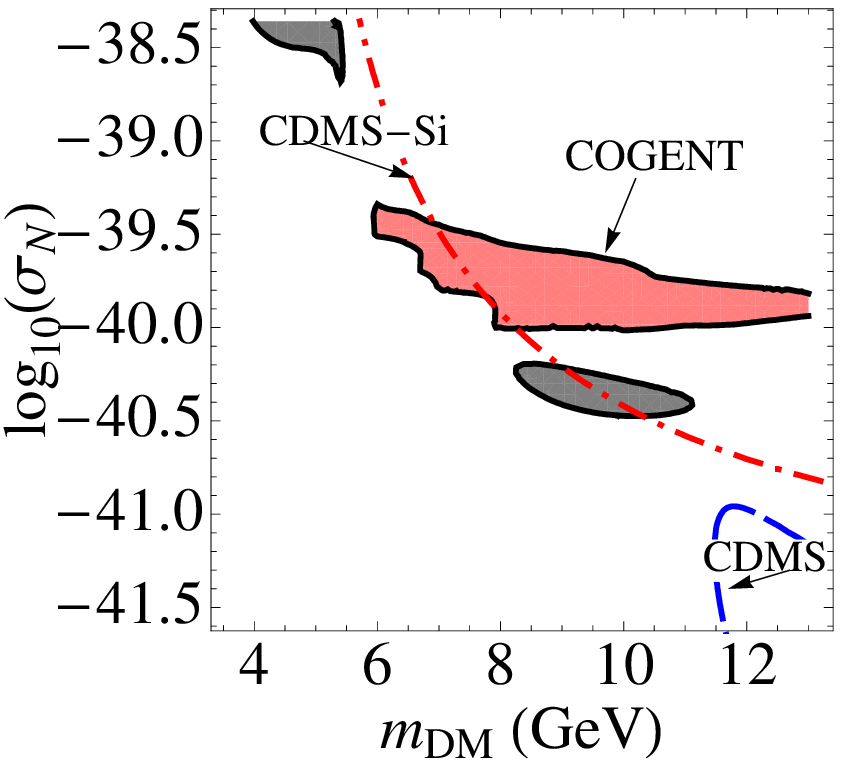}
\caption{{\em Left panel:} The DAMA region and XENON10 constraint are the same as in the right panel of Fig.~\ref{fig:noLeff} but, for illustration, with the effects of removing the lowest bin of data from the DAMA region (gray), or with the effects of increasing the errors in those bins by a factor of two (purple and blue
for 90\% and 99\%).  We see that the DAMA region is shifted to smaller masses and becomes more consistent with the XENON10 constraint.
%Same as in the previous figure, but with the effects of the halo model on the consistency of the CoGeNT window with DAMA and the null experiments.  $v_0 = 270$ km/s and $v_{\rm esc} = 650 \mbox{ km/s}$.  We see that the DAMA region is shifted to smaller masses and becomes more consistent with the CoGeNT region. 
{\em Right panel:} The same in in the left frame, but with dark matter couplings only to neutrons, $f_n = 1$ and $f_p = 0$.
%xxx-liam
 The constraints from CDMS silicon relative to the
DAMA region are weaker than for $f_n=f_p$.
%xxx
Only the 90\% confidence limit for DAMA removing the
lowest bin is shown.}
\label{fig:halo}
\end{figure}

To further investigate the source of the remaining tension, we examine the effects of both removing the lowest bin from the DAMA data sample, and making the errors in those bins a factor of two larger.  In the left frame of Fig.~\ref{fig:halo}, we have modified
the DAMA constraint by showing the region when lowest bin is neglected
altogether (grey, 90\% and 99\% constraints), as well as the region
when the errors are doubled on the lowest two bins (purple, blue, for
90\% and 99\%, respectively).  We can see that most of the remaining tension is 
%xxx-liam
%removed so that we can see that most of the inconsistency seems to stem from the lowest one to two bins in the DAMA data.
removed: much of the inconsistency seems to stem from the lowest one to two bins in the DAMA data.  
%xxx

In the results shown 
%xxx-liamin Fig.~\ref{fig:noLeff},
so far,
%xxx
we have assumed that the couplings of the dark matter to protons are equal to those to neutrons.  If the dark matter couples to neutrons more strongly than to protons, we can further relax the constraints of the CDMS silicon analysis.  For example, taking $f_n = 1$, $f_p = 0$ we find the results shown in the right frame of Fig.~\ref{fig:halo}.

%{\bf BEGIN 1}

%\begin{figure}[htp]
%\centering
%\includegraphics[width=3.5in]{banana-4-fpfn.eps}
%\caption{}
%\label{fig:fpfn}
%\end{figure}

%{\bf END 1}

\begin{table}[tbhp]
\centering
\begin{tabular}{|c|c|c|c|c|c|} \hline
Experiment & Target & Exposure (kg-d) & Threshold  & Ref \\ \hline\hline
%CDMS-SUF & Ge & 65.8 & 5 \mbox{ keV} & \cite{CDMSI} $$ \\ 
%&                 Si   & 6.58 & 5 \mbox{ keV} & \\ \hline
CDMS-II & Ge & 
%xxx-liam
612.
%xxx
 & 10 \mbox{ keV} & \cite{CDMSII}\\ 
& Si & 
%xxx-liam
83.3 
%xxx
& 7 \mbox{ keV} & 
%xxx-liam
%\cite{CDMSIISi}
\cite{CDMSIISi,CDMSIISi2}
%xxx
 \\ \hline
XENON10 & Xe & 
%xxx-liam
316 
%xxx
& 2 \mbox{ keV} \mbox{ with} $L_{eff} = 0.19$ &  \cite{XENONinelastic}\\ \hline
%CRESST-I & Al$_2$O$_3$ & 1.51 &  $0.6 \mbox{ keV}$ &  \cite{CRESST} \\ \hline
%TEXONO & Ge & 0.338 & $0.2 \mbox{ keVee}$ & \cite{TEXONO}\\ \hline
 \hline

\end{tabular}
\caption{\label{ExpFacts} Relevant features of the null experiments used in our analysis.
%xxx-liam
Efficiencies and cut acceptances are taken from the references above.
%xxx
}
\end{table}

\section{Models of Low Mass Dark Matter}

We begin this section by carrying out a relatively model independent operator analysis of the low mass window relevant for CoGeNT, DAMA and CDMS.  We then turn to
discussing two classes of models: those arising from models of Asymmetric Dark Matter (ADM), and from supersymmetry.\footnote{Singlet scalars \cite{singlet}, Mirror models \cite{mirror} and ``WIMPless'' models \cite{wimpless} also give rise to light WIMPs, though we do not discuss those models here.  We refer the reader to these references for details.}

%xxx-liam
%\section{Operator Analysis}
\subsection{Operator Analysis}
\label{MI}
%xxx
The operators connecting dark sector Dirac fermions $\chi$ or complex scalars $\phi$ to visible sector fermions $f$ are
\begin{eqnarray}
{\cal L}_{fS}& = & \frac{G_{f}}{\sqrt{2}} \bar{\chi} \chi \bar{f} f \\
{\cal L}_{fV}& = & \frac{G_{f}}{\sqrt{2}} \bar{\chi} \gamma^\mu \chi \bar{f} \gamma_\mu f \\
{\cal L}_{sS}& = & \frac{F_{f}}{\sqrt{2}} \phi^* \phi \bar{f} f \\
{\cal L}_{sV}& = & \frac{F_{f}}{\sqrt{2}} \phi^* \partial_\mu \phi \bar{f} \gamma^\mu f.
\end{eqnarray}

\subsubsection{Fermionic dark matter}

For fermionic dark matter, the scattering cross-section obtained from these operators is given by:
\be\label{scattering-fs}
\sigma=\frac{4 a}{\pi}\frac{m_{\rm DM}^2m_N^2}{(m_{\rm DM}+m_N)^2}\left(Z f_p+(A-Z)f_n\right)^2,
\ee
where $A$ and $Z$ are the atomic mass and atomic number of the target nuclei, and $a$ is a number dependent on whether the fermion is Dirac of Majorana.  For Majorana fermions $a = 1$, while for Dirac fermions, $a = 1/4$.  The effective couplings to protons and neutrons, $f_{p,n}$, can be written in terms of the WIMP's couplings to quarks, and depend both on the spin of the DM particle and the mediator.   In terms of the dark matter's effective couplings, we can write
\be \label{fpn}
f_{p,n}=\sum_{q=u,d,s} \frac{G_{q}}{\sqrt{2}} f^{(p,n)}_{Tq}\frac{m_{p,n}}{m_q}+\frac{2}{27}f^{(p,n)}_{TG}\sum_{q=c,b,t}  \frac{G_{q}}{\sqrt{2}} \frac{m_{p,n}}{m_q},
\ee
where $G_q$ denotes the dark matter's effective coupling to a given quark species. The first term reflects scattering with light quarks, while the second term accounts for interactions with gluons through a heavy quark loop. The values of $f^{(p,n)}_{T_q}$ are proportional to the matrix element, $\langle \bar q q\rangle$, of quarks in a nucleon. In our numerical calculations, we use values for these quantities based on recent lattice QCD results~\cite{lattice,lattice2}. If the dark matter scatters with nuclei through a scalar interaction, both of the terms in Eq.~\ref{fpn} typically yield sizable contributions to the cross section. If, on the other hand, the dark matter scatters through a vector interaction, then heavy-quark loop contribution is negligible, and the nucleon level couplings simplify to 
\be
f_{p} = 2 \frac{G_{u}}{\sqrt{2}}+\frac{G_{d}}{\sqrt{2}}, \,\,\,\,\,\,\,\,\,\,\,\,\, f_{n} = \frac{G_{u}}{\sqrt{2}}+2\frac{G_{d}}{\sqrt{2}}.
\ee 

The effective operator relates to the parameters in the underlying Lagrangian through
\be
\frac{G_{f}}{\sqrt{2}} = \frac{g_{\rm DM} g_f}{M^2_{\psi}},
\label{fermicoupling}
\ee
where $\psi$ denotes the mediator, and $g_{\rm DM}$, $g_{f}$ denote the mediator's couplings to the dark matter and standard model fermion, respectively. 

The annihilation cross-sections for Dirac fermions $\bar{\chi} \chi \rightarrow \bar{f} f$, through scalar, pseudoscalar and vector mediators, are
\begin{eqnarray}
\sigma_{ann}^{fS} v & = & \frac{v^2 m_{DM}^2}{16 \pi} \sum_f G_f^2 c_f (1-m_f^2/m_{DM}^2)^{3/2} \\
\sigma_{ann}^{fP} v & = & \frac{m_{DM}^2}{4 \pi} \sum_f G_f^2 c_f \left[(1-m_f^2/m_{DM}^2)^{1/2} - \frac{1-m_f^2/2m_{DM}^2}{4(1-m_f^2/m_{DM}^2)^{1/2}} \right]\\
 \sigma_{ann}^{fV} v & = &  \frac{m_{DM}^2}{4 \pi} \sum_f G_f^2 c_f (1-m_f^2/m_{DM}^2)^{1/2} \left[\left(2 +\frac{m_f^2}{m_{DM}^2}\right)+\left(\frac{8-4\frac{m_f^2}{m_{DM}^2}+5 \frac{m_f^4}{m_{DM}^4}}{24(1-\frac{m_f^2}{m_{DM}^2})}\right) v^2  \right] ,
 \end{eqnarray}
 where $c_f$ is a color factor.

\subsubsection{Scalar Dark Matter}
There are related expressions for scalar dark matter scattering and annihilation cross-sections:
\be\label{scattering-ss}
\sigma=\frac{a}{\pi}\frac{m_N^2}{(m_{\rm DM}+m_N)^2}\left(Z f_p+(A-Z)f_n\right)^2,
\ee
with
\be \label{fpn2}
f_{p,n}=\sum_{q=u,d,s} \frac{F_{q}}{\sqrt{2}} f^{(p,n)}_{Tq}\frac{m_{p,n}}{m_q}+\frac{2}{27}f^{(p,n)}_{TG}\sum_{q=c,b,t}  \frac{F_{q}}{\sqrt{2}} \frac{m_{p,n}}{m_q},
\ee
for scalar mediators, and 
\be
f_{p} = 2 \frac{F_{u}}{\sqrt{2}}+\frac{F_{d}}{\sqrt{2}}, \,\,\,\,\,\,\,\,\,\,\,\,\, f_{n} = \frac{F_{u}}{\sqrt{2}}+2\frac{F_{d}}{\sqrt{2}},
\ee 
for vector mediators.  $a=1$ for real scalars, and $a=1/4$ for complex scalars.

For scalar mediators, the effective coupling is related to the underlying parameters in the Lagrangian through
\be
\frac{F_f}{\sqrt{2}} = \frac{g_{\rm DM} y_f \langle H \rangle}{M_h^2},
\ee
where $v$ is the Higgs vev, $v/\sqrt{2} = \langle H \rangle = 174 \mbox{ GeV}$, and $M_h$ is its mass, derived from a term in the Lagrangian  
\be
{\cal L} = g_{\rm DM} H^\dagger H S^\dagger S + y_f H \bar{f} f.
\ee
For vector mediators (and complex scalars), the effective coupling is related to the underlying parameters through
\be
\frac{F_f}{\sqrt{2}} = \frac{g_{\rm DM} g_f m_{\rm DM}}{M_\psi^2}.
\ee 

The corresponding annihilation cross-sections for a complex scalar $\bar{\phi} \phi \rightarrow \bar{f} f$, through real scalar and vector mediators, are
\begin{eqnarray}
\sigma_{ann}^{sS} v & = &  \frac{1}{8 \pi} \sum_f F_f^2 c_f (1-m_f^2/m_{DM}^2)^{1/2}\left[\left(1-\frac{m_f^2}{m_{DM}^2}\right)+\left(\frac{3 m_f^2}{8 m_{DM}^2}\right) v^2 \right] \\
 \sigma_{ann}^{sV} v & = &  \frac{v^2}{12 \pi} \sum_f F_f^2 c_f (1-m_f^2/m_{DM}^2)^{1/2}\left(1+\frac{m_f^2}{2 m_{DM}^2}\right).
\end{eqnarray}

\subsubsection{CoGeNT and Thermal WIMPs}

\begin{figure}[t]
\centering\leavevmode
\includegraphics[width=3.3in,angle=0]{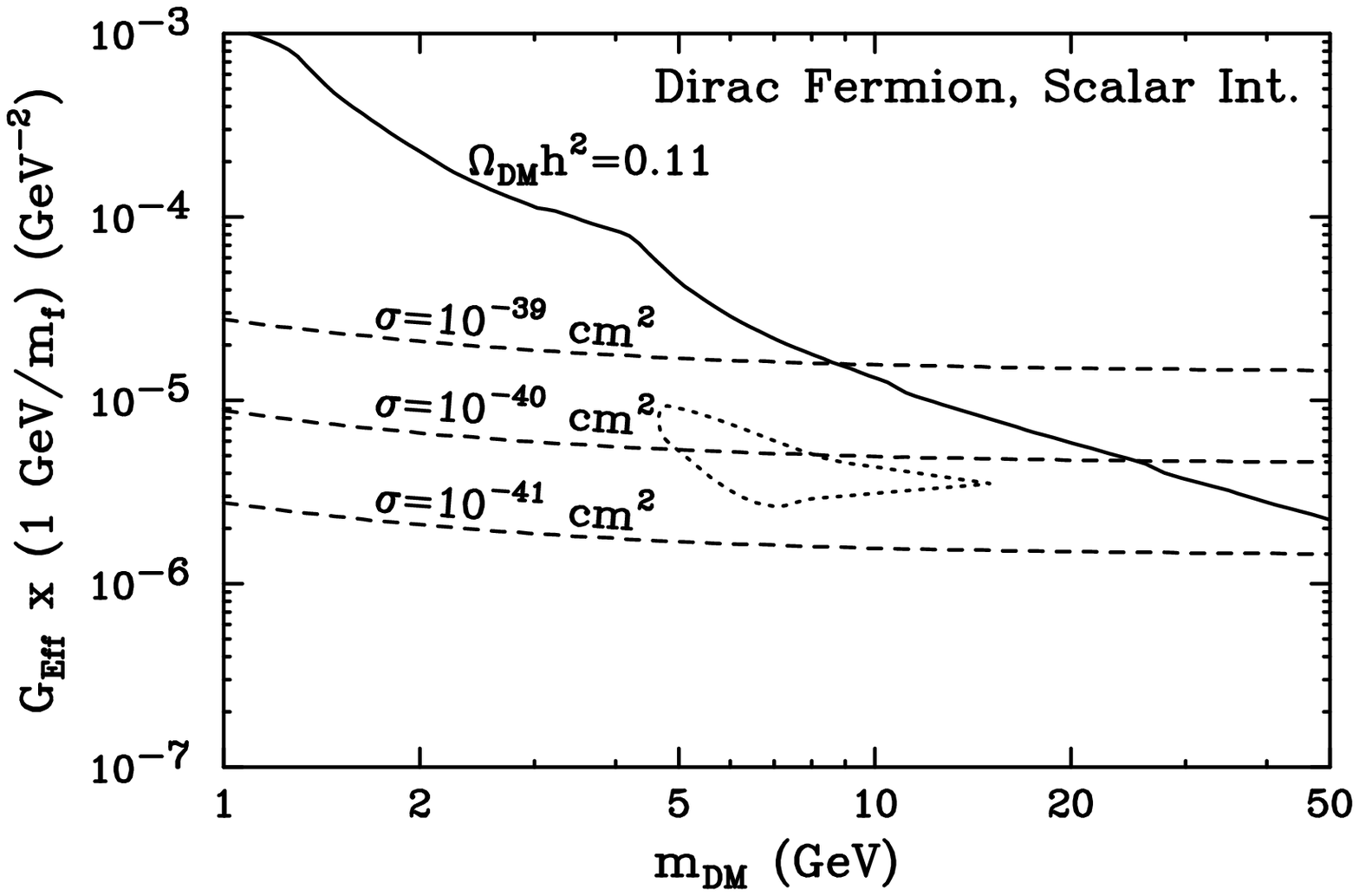}
\includegraphics[width=3.3in,angle=0]{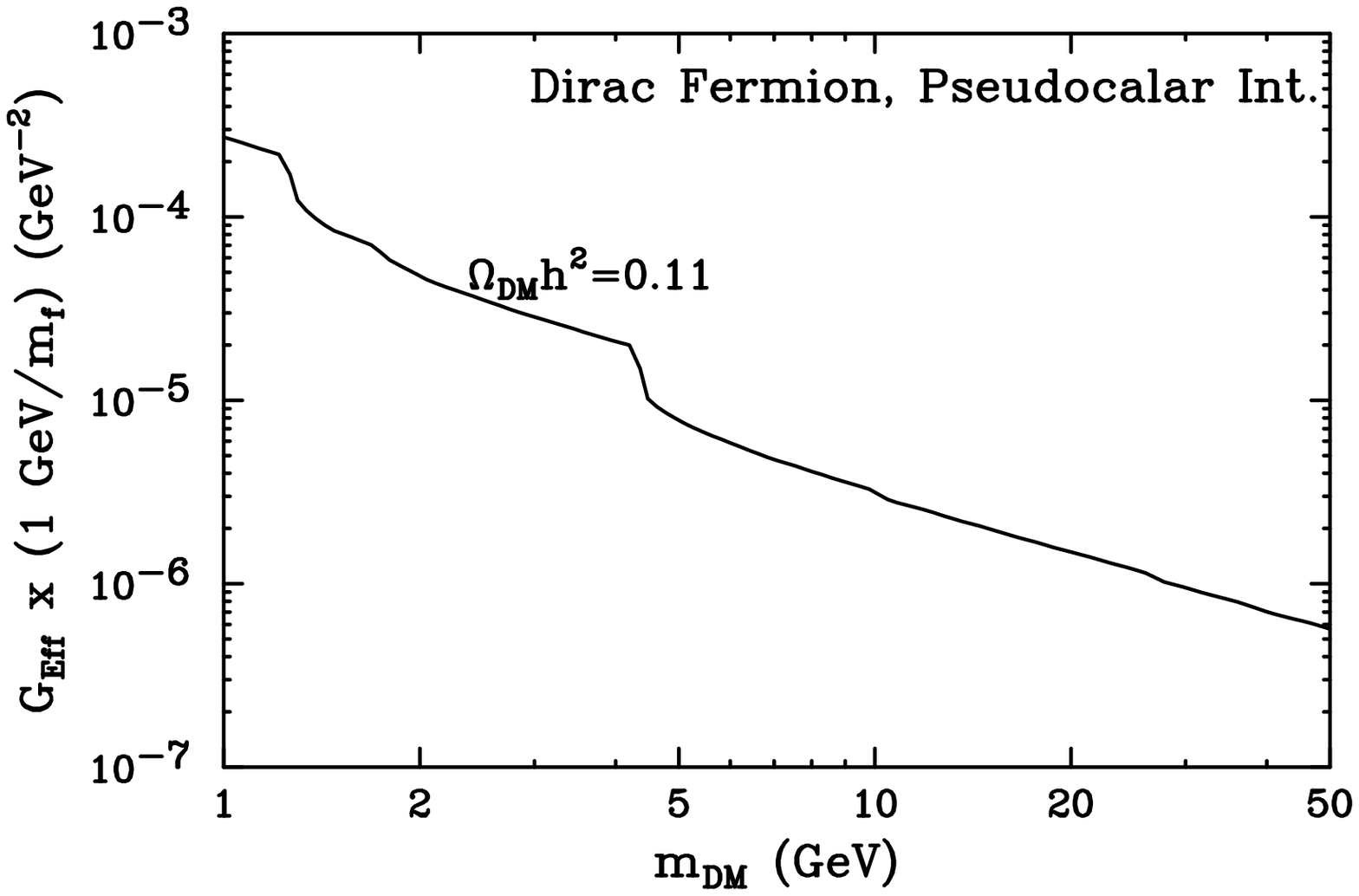}\\
\includegraphics[width=3.3in,angle=0]{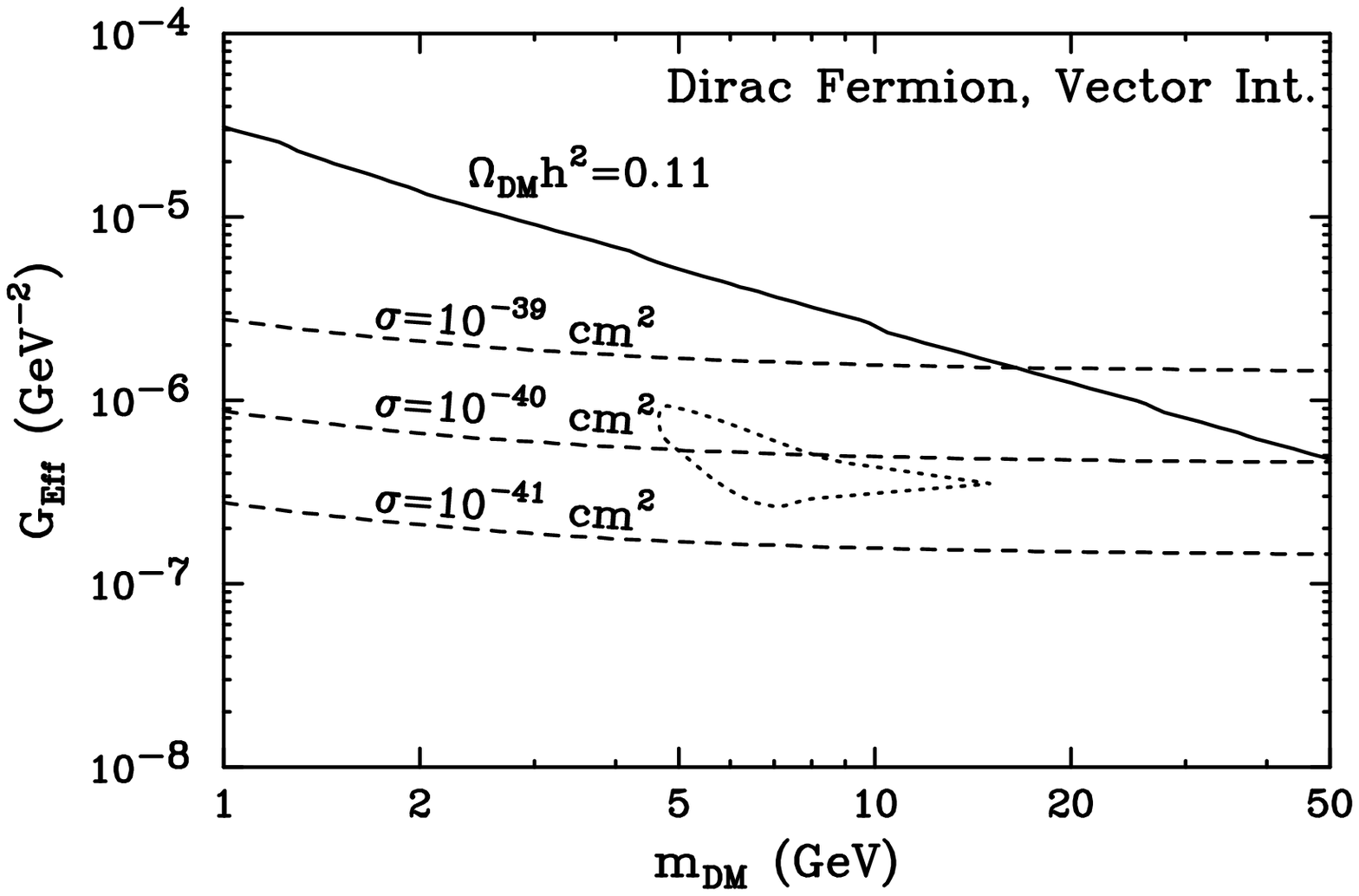}\\
\includegraphics[width=3.3in,angle=0]{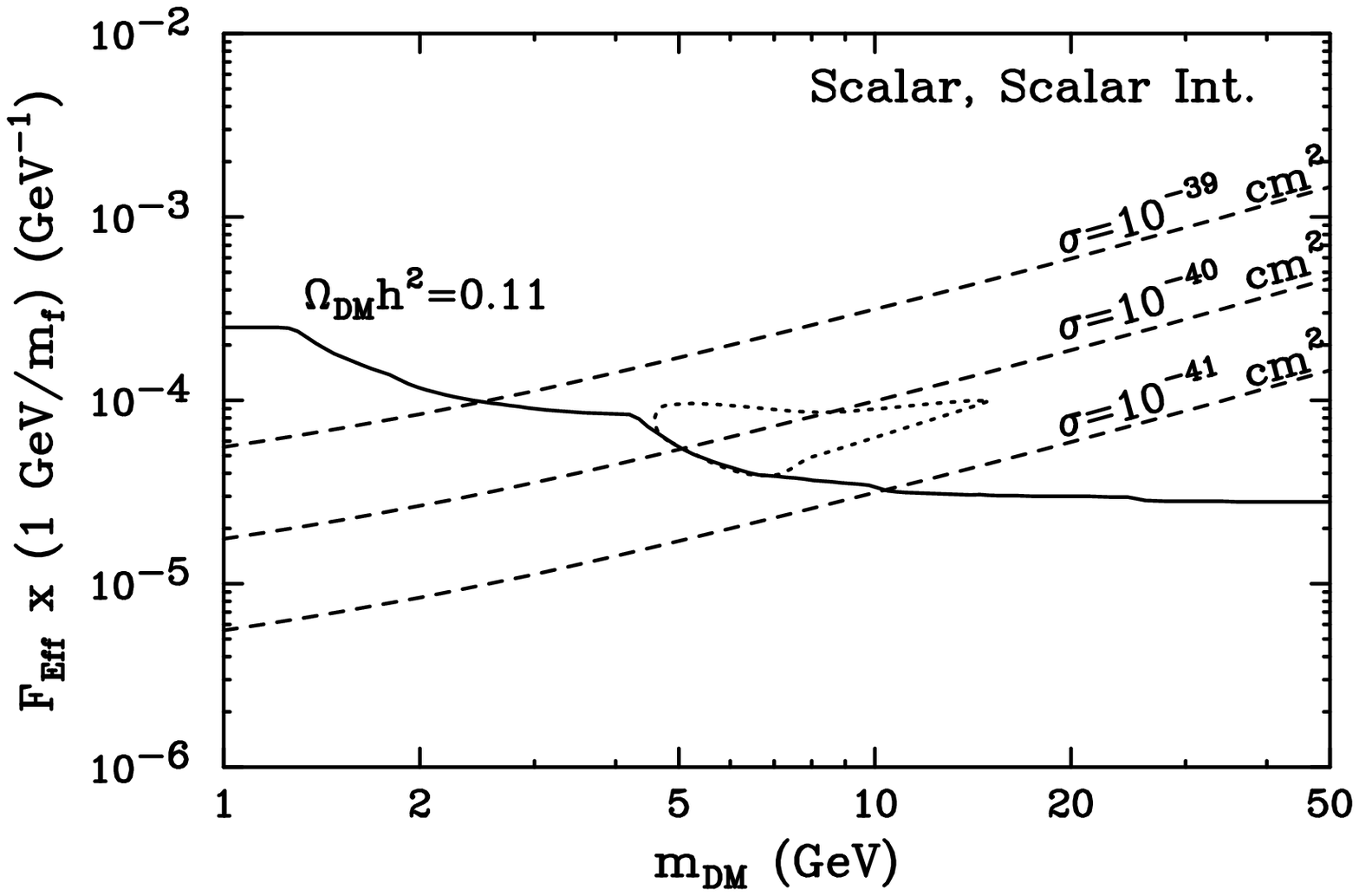}
\includegraphics[width=3.3in,angle=0]{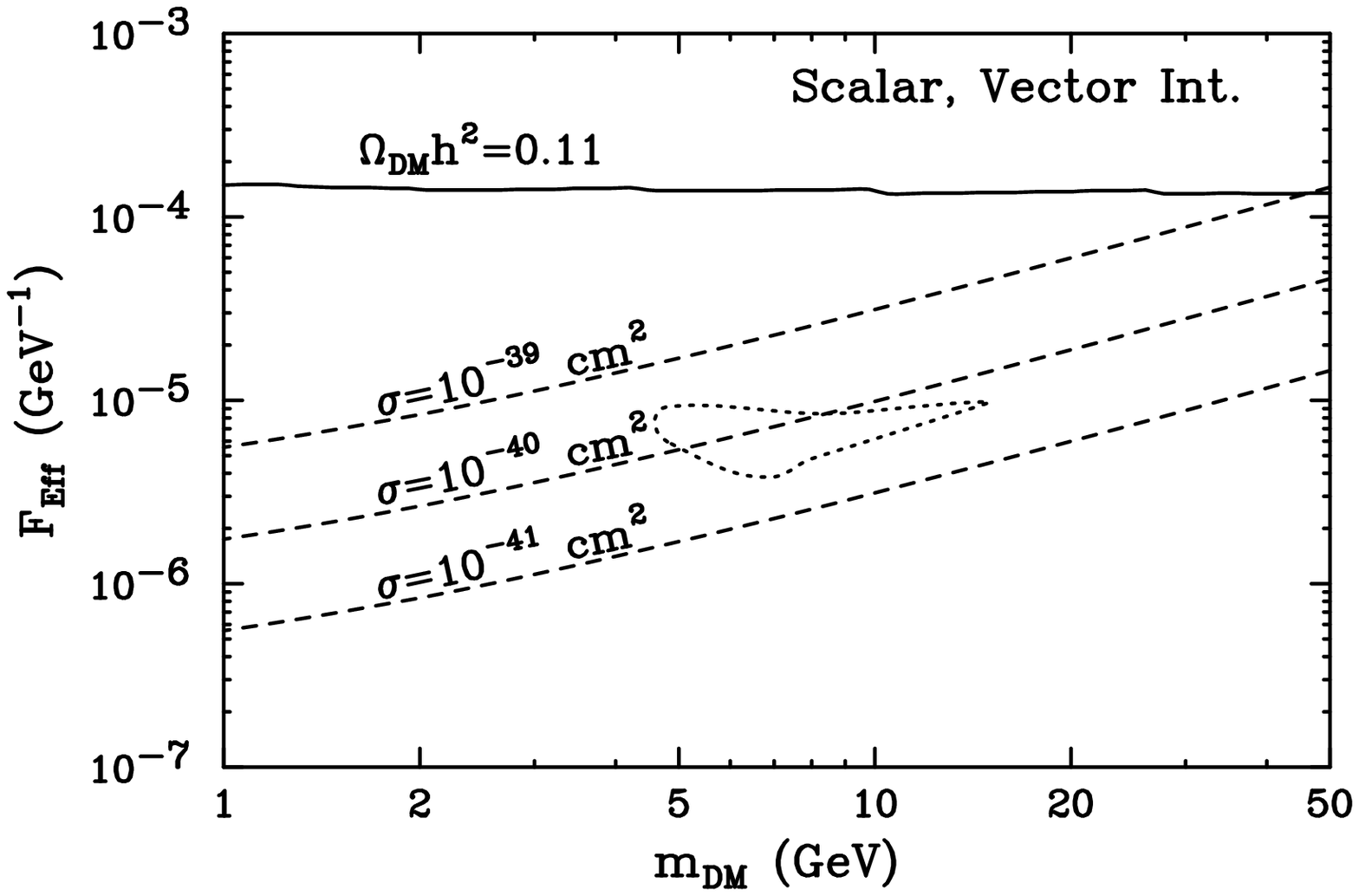}
\caption{The effective dark matter-fermion couplings required to generate a given spin-independent elastic scattering cross section with nucleons (dashed lines), and to produce a thermal abundance consistent with the observed dark matter density (solid lines). In each frame, the dotted region represents the approximate range of couplings and masses that leads to a signal consistent with that reported by CoGeNT. The various frames show results for a fermionic dark matter particle with scalar interactions (upper left), pseudoscalar interactions (upper right), vector interactions (center), or a scalar dark matter particle with scalar interactions (lower left), or vector interactions (lower right). In the case of a scalar dark matter particle with approximately universal scalar couplings to Standard Model fermions, the coupling size required to generate the excesses observed by CoGeNT and DAMA is also approximately the size required to thermally generate the observed density of dark matter in the early universe.}
\label{summary}
\end{figure}

Beginning with the case of a fermionic WIMP, we can consider either scalar or vector interactions with quarks to produce a spin-independent elastic scattering with nuclei. Following Ref.~\cite{modelindependent}, we show results for each of these cases in Fig.~\ref{summary}. In the upper left frame, we have assumed that the dark matter's effective scalar couplings to fermions are proportional to the mass of the fermion, $G_f \propto m_f$, but are otherwise universal. To produce an elastic scattering cross section capable of generating the excesses observed by CoGeNT and DAMA (shown as a dotted region), we need an effective coupling of $G_{\rm Eff} \times (1\,{\rm GeV}/m_f) \sim 2 \times 10^{-6}$ GeV$^{-2}$. If this is the only interaction experienced by the WIMP, however, it will be significantly overproduced in the early universe, resulting in a thermal relic abundance in excess of the measured abundance of dark matter. The solid lines in each frame of Fig.~\ref{summary} denote the effective coupling strength that leads to the desired thermal dark matter abundance, assuming no other interactions further deplete the relic density. For details pertaining to the dark matter's annihilation cross section and the thermal relic abundance calculation, see Ref.~\cite{modelindependent}.

This apparent conflict between the couplings required to produce the desired elastic scattering rate and the measured dark matter abundance can easily be resolved, however, if the fermionic WIMP also annihilates through other process, such as through pseudoscalar or axial interactions (which do not, in turn, generate a contribution to the spin-independent elastic scattering cross section). In the upper right frame of Fig.~\ref{summary}, we show the effective pseudoscalar coupling that leads to the desired dark matter abundance. A 9 GeV WIMP, for example, with effective scalar {\it and} pseudoscalar couplings of $G_{\rm Eff} \sim 3 \times 10^{-6}$ GeV$^{-2}  (m_f/1\,{\rm GeV})$ would satisfy both the elastic scattering and relic density requirements. Alternatively, one could also consider effective couplings that are not universal across all fermion species. If the dark matter couples preferentially to leptons, for example, its relic abundance could be reduced without increasing the elastic scattering cross section.

As a simple example, we could consider a light fermionic dark matter particle that couples to quarks through the exchange of the Standard Model Higgs boson. This scenario, however, is unlikely to be able to generate the excesses reported by CoGeNT and DAMA. In particular, even if we take the dark matter's couplings to the Higgs to be of order unity, the effectively couplings to quarks would only be $G_{\rm Eff} \sim 4\times 10^{-7}$ GeV$^{-2} (m_f/1\,{\rm GeV})$, or approximately an order of magnitude smaller than is needed to generate the observed number of events. From this example, we see that the mediator must either be considerably lighter than the Standard Model Higgs collider limits, or must have coupling to quarks that are larger than those of the Standard Model Higgs. 

If the dark matter is a Dirac fermion, vector couplings can also generate a spin-independent elastic scattering cross section. In the center frame of Fig.~\ref{summary}, we show the effective couplings required in this case.  Here, we have assumed that $G_{\rm Eff}$ is the same for all species of Standard Model fermions. Once again, we find that the couplings required to generate the desired elastic scattering rate are not sufficient to yield the observed dark matter abundance, and other interactions are required to avoid the overproduction of dark matter in the early universe. 

If instead of a fermion, the WIMP is a scalar particle, we again can rely on either scalar or vector interactions as the source of the CoGeNT and DAMA excesses. In the case of scalar interactions, we find that the effective coupling required to generate the elastic scattering cross section implied by CoGeNT and DAMA also leads to a thermal relic abundance consistent with the observed dark matter density (see the lower left frame of Fig.~\ref{summary}). So, in this simplest case we have considered (a scalar dark matter particle with approximately universal scalar couplings to Standard Model fermions), we find that if we set the couplings to the size required to generate the observed density of dark matter, then we also predict an elastic scattering rate that can easily account for the CoGeNT and DAMA excesses. A scalar WIMP with vector interactions, on the other hand, requires another interaction to avoid being overproduced in the early universe (see the lower right frame of Fig.~\ref{summary}).

\subsection{Asymmetric Dark Matter}
\label{adm}

In most dark matter models, the abundance of dark matter and the baryon asymmetry of the universe are unrelated to each other.  The dark matter density is generally determined by its mass and self-annihilation cross section, which determine the temperature at which is thermally freezes out in the early universe.  The baryon asymmetry, in contrast, is set by CP violating phases, and by the strength of the electroweak phase transition. Within this paradigm, the densities of dark matter and of baryons in the universe have nothing to do with one another. But despite this expectation, observations have revealed that the cosmological densities of these two sectors are similar:
\begin{equation}
\frac{\rho_{DM}}{\rho_b} \approx 5.
\label{coin}
\end{equation}
Within the context of most thermal dark matter models, one must view this similarity as a coincidence (unless anthropic arguments are invoked).

Motivated by this apparent coincidence, models in which the dark matter abundance is tied to the baryon asymmetry have been developed~\cite{ADMmodels,ADM}. In ADM models, a mechanism (in some cases a higher dimension operator) enforces a relationship between the dark matter and baryon chemical potentials:
\begin{equation}
c_1(n_X - n_{\bar{X}}) = n_b - n_{\bar{b}},
\end{equation}
where $c_1$ is a number ${\cal O}(1)$ whose precise value depends on the operator transferring the asymmetry.  

Once the symmetric thermal abundance, $n_X + n_{\bar{X}}$, has annihilated away, we are left with a relationship between the asymmetric relic dark matter density and the relic baryon number density, implying
\begin{equation}
m_{DM} = 5 c_1 m_p,
\end{equation}
which combined with Eq.~(\ref{coin}) requires that the dark matter's mass is roughly within the range of $\sim$1-10 GeV.

In Ref.~\cite{ADM}, a class of models was discussed in detail that fits naturally within the CoGeNT, DAMA and CDMS windows.  In one such model (a supersymmetric example), the dark matter and baryon chemical potentials are related to each other through a superpotential interaction
\begin{equation}
W = \frac{\bar{X}^2 LH}{M}.
\end{equation}
This operator transfers a lepton asymmetry to the dark matter and predicts a dark matter mass of approximately $m_{X} \sim $11 GeV.
%\begin{equation}
%m_{X} \approx 11 \mbox{ GeV}
%\end{equation}
Other classes of models have also been discussed in which the dark matter is related to the baryon density through the out-of-equilibrium decay of a new heavy particle to both sectors.

ADM thus predicts dark matter with a mass near or within the window implied by CoGeNT and DAMA. In order to provide an explanation for these excesses, however, such a dark matter candidate must also possess an appropriately large elastic scattering cross section. The magnitude of this cross section is determined not only by the properties of the dark matter particle itself, however, but also on the mass and couplings of the mediator, which makes it a somewhat more model dependent quantity. As one possibility (as discussed in Ref.~\cite{ADM}), a simple $Z'$ mediator with mass around the TeV scale gives rise to a scattering cross section of the magnitude observed by CoGeNT and DAMA.  Alternatively, a singlet scalar that mixes with the Higgs may also give rise to an appropriately sized elastic scattering cross section.  This singlet may be responsible for the mass of the singlet dark matter through an operator
\begin{equation}
W_M = \lambda S \bar{X} X,
\end{equation}
and a mixing with the Higgs through the usual NMSSM term
\begin{equation}
W_n = \zeta S H_u H_d
\end{equation}
gives rise to the appropriate size scattering cross section for CoGeNT and DAMA~\cite{multicomp}:
\begin{eqnarray}
%xxx-liam 
%I changed m_r to \mu_n and \sigma_n to \sigma_N to be consistent with notation in section1
\sigma_N & \approx & \frac{\mu_n^2}{\pi} N_n^2 \left(\frac{\lambda \zeta v_u \langle S \rangle}{m_{h^0}^2} \right)^2 \frac{1}{m_S^4} \nonumber \\
& \approx & 3 \times 10^{-41} \mbox{ cm}^2 \left(\frac{N_n}{0.1}\right)^2  \left(\frac{\lambda}{0.2}\right)^2 \left(\frac{\zeta}{10^{-2}}\right)^2 \left(\frac{\langle S \rangle}{20 \mbox{ GeV}}\right)^2 \left(\frac{100 \mbox{ GeV}}{m_{h^0}}\right)^4 \left(\frac{10 \mbox{ GeV}}{m_S}\right)^4,
\end{eqnarray}
where $N_n$ arises from the effective coupling of the Higgs to the target nucleus.\footnote{Note that the relic abundance arguments described in Sec.~\ref{MI} do not apply in the case of asymmetric dark matter. Furthermore, the indirect detection prospects described in Sec.~\ref{indirect} do not apply to this class of dark matter candidates.}

\subsection{(Non-Minimal) Neutralino Dark Matter}

Within the context of the minimal supersymmetric standard model (MSSM), light neutralino dark matter can coherently scatter with nuclei through scalar Higgs exchange (and to a lesser extent, through squark exchange), while annihilating fairly efficiently through pseudoscalar Higgs exchange (which, unlike scalar exchange, is not $s$-wave suppressed). The composition and couplings of a neutralino lighter than $m_Z/2$ is constrained by the LEP measurements of the invisible $Z$ branching fraction, which require $\Gamma_{Z \rightarrow \chi^0 \chi^0} < 3$ MeV (at $2\sigma$ C.L.)~\cite{zwidth}. This approximately translates to a bound of $|N^2_{13} - N^2_{14}| \lsim 0.13$ at $2 \sigma$, where $N^2_{13}$ and $N^2_{14}$ are the up-type and down-type higgsino fractions of the lightest neutralino. If we maximize the neutralino-neutralino-Higgs couplings by saturating this bound, and impose the constraints from the Tevatron for Higgs boson production at large $\tan \beta$, it has been shown that the largest scattering cross-section that can be achieved is $\sim 2 \times 10^{-41} \mbox{ cm}^2$ \cite{MSSMnew}. This is near to the CoGeNT and DAMA regions \cite{italians,lightLSP}, but certainly on the lower edge.   

To obtain larger effective couplings (and thus larger elastic scattering and annihilation cross sections), one could consider interactions mediated by Higgs bosons which are lighter than those allowed within the MSSM. Within the context of the next-to-minimal supersymmetric standard model (NMSSM), for example, light ($\sim$1-100 GeV) scalar and pseudoscalar Higgs bosons can exist, while not conflicting with existing collider constraints, as a result of their reduced couplings to Standard Model fields. In particular, if the lightest scalar and pseudoscalar Higgs bosons contain a significant admixture of the a Higgs singlet, such particles can be much lighter than the MSSM Higgs limits, increasing the corresponding effective couplings of the lightest neutralino~\cite{gunion,ferrer}.  

As an example, we could consider a 5-10 GeV neutralino which is mostly bino, but with a $\sim$$5$\% higgsino admixture. If such a particle annihilates through the exchange of a $\lsim 70$ GeV mixed singlet-MSSM-like pseudoscalar Higgs, and if $\tan \beta$ is relatively large ($\sim$$30$-$50$), then we find that this process can yield the desired relic abundance. Furthermore, if a down-type scalar Higgs ($H_1$) were also of a comparable mass, and comparably mixed with the Higgs singlet, it would mediate an elastic scattering cross section consistent with the rate observed by CoGeNT and DAMA. This example serves as an illustration of the more general conclusion that the range of masses and scattering cross sections implied by CoGeNT and DAMA requires a particle or particles that are either relatively light or possess relatively large couplings to mediate the interactions of the dark matter.

\section{Implications For Indirect Detection}
\label{indirect}

Regions with high densities of dark matter are in many cases predicted to produce potentially observable fluxes of dark matter annihilation products. As the annihilation rate of dark matter scales with the square of the inverse mass of the particle, the prospects for indirect detection are particularly encouraging for dark matter masses as light as we are considering in this paper. In this section, we discuss some of the implications of the CoGeNT and DAMA excesses for indirect dark matter searches using neutrinos, gamma rays, and charged cosmic rays. We also comment on the possibility that white dwarf stars could be used as a probe of dark matter in models capable of generating the observed excesses.

\subsection{Neutrinos From Dark Matter Annihilations In the Sun}

If dark matter particles are present in the vicinity of the solar system, they will scatter elastically with and become captured in the Sun at a rate approximately given by~\cite{capture}
\begin{eqnarray}
C^{\odot} \approx 1.3 \times 10^{24} \, \mathrm{s}^{-1} 
\left( \frac{\rho_{0}}{0.3\, \mathrm{GeV}/\mathrm{cm}^3} \right) 
\left( \frac{270\, \mathrm{km/s}}{\bar{v}} \right)  
\left( \frac{10 \, \mathrm{GeV}}{m_{\rm DM}} \right) 
\sum_A F_A  \bigg(\frac{\sigma_{\mathrm{A}}} {A^2 \times 10^{-40}\, {\rm cm}^2}\bigg)  S(m_{\rm DM}/m_{{\rm A}}),
\label{capture}
\end{eqnarray}
where $\rho_{0}$ is the local dark matter density, $\bar{v}$ is the local root-mean-square velocity of halo dark matter particles, and $m_{\rm DM}$ is the dark matter mass. The sum is over mass number of the chemical species present in the Sun, and the quantity $F_A$ contains information pertaining to the solar abundances, dynamical factors, and form factor suppression of each element $(F_{\rm H}=1, F_{\rm He}\approx 1.1$, etc.). The quantity $S$ is a kinetic suppression factor given by
\begin{equation}
S(x)=\bigg[\frac{A(x)^{3/2}}{(1+A(x)^{3/2})}\bigg]^{2/3},
\end{equation}
where
\begin{equation}
A(x)=\frac{3x}{2(x-1)^2}\bigg(\frac{\langle v_{\rm esc}
\rangle}{\bar{v}}\bigg)^2.
\end{equation}
%
%For WIMPs much heavier than the nuclei they are scattering with, this suppression can be considerable. For WIMPs in the 7--11 GeV range, however, $S$ is near unity.%

The evolution of the number of dark matter particles in the Sun, $N$, is described by 
\begin{equation}
\dot{N} = C^{\odot} - A^{\odot} N^2 - E^{\odot} N,
\end{equation}
where $C^{\odot}$ is the capture rate, $A^{\odot}$ is the 
annihilation cross section times the relative WIMP velocity per unit volume, and $E^{\odot}$ is the inverse time for a WIMP to exit the Sun via evaporation (which is negligible for dark matter particles heavier than approximately 3-4 GeV~\cite{evaporation,equ1}). $A^{\odot}$ can be approximated by
\begin{equation}
A^{\odot} = \frac{\langle \sigma v \rangle}{V_{\mathrm{eff}}}, 
\end{equation}
where $V_{\mathrm{eff}}$ is the effective volume of the core
of the Sun determined roughly by matching the core temperature with 
the gravitational potential energy of a single WIMP at the core
radius.  This was found in Refs.~\cite{equ1,equ2} to be
\begin{equation}
V_{\rm eff} = 5.7 \times 10^{30} \, \mathrm{cm}^3 
\left( \frac{1 \, \mathrm{GeV}}{m_{\rm DM}} \right)^{3/2} \;.
\end{equation}
Neglecting evaporation, the present WIMP annihilation rate in the Sun is given by
\begin{equation} 
\Gamma = \frac{1}{2} A^{\odot} N^2 = \frac{1}{2} \, C^{\odot} \, 
\tanh^2 \left( t_{\odot} \sqrt{C^{\odot} A^{\odot}} \right) \;, 
\label{rate:noevap}
\end{equation}
where $t_{\odot} \approx 4.5$ billion years is the age of the solar system.
For a 5-10 GeV WIMP, capture-annihilation equilibrium is reached so long as the WIMP's annihilation cross section is larger than $\sigma v \gsim 10^{-30}$ cm$^3$/s. Once equilibrium is reached, the final annihilation rate (and corresponding neutrino flux and event rate) has no further dependence on the dark matter particle's annihilation cross section.

\begin{figure}[t]
%\begin{center}
 \includegraphics[width=0.5\textwidth,angle=0]{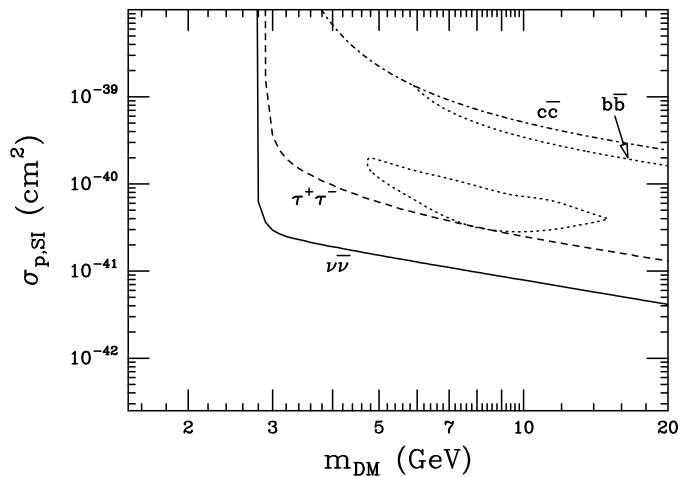}
%\resizebox{11.2cm}{!}{\includegraphics{l.ps}}
%\resizebox{11.2cm}{!}{\includegraphics{limitSI2.ps}}
\caption{The upper limit on a light WIMP's spin-independent elastic scattering cross section with nuclei from the Super Kamiokande experiment, for various choices of the dominant annihilation channel. Also shown is the region of the plane which provides a good fit to the excess observed by CoGeNT. From this figure, we conclude that Super Kamiokande excludes a dark matter particle which annihilates entirely to neutrinos or taus as a source of the CoGeNT excess. This figure was adapted from Ref.~\cite{Hooper:2008cf}.}
\label{superk}
%\end{center}
\end{figure}

As the dark matter particles annihilate, they can generate neutrinos through a wide range of channels. Annihilations to bottom quarks, charm quarks, and tau leptons each generate neutrinos in their subsequent decays. In some models, dark matter particles can also annihilate directly to neutrino pairs. Once produced, neutrinos travel to the Earth where they can be detected. For the mass range of interest here, the strongest constraints on the capture and annihilation rate in the Sun come from the Super-Kamiokande experiment~\cite{superk}. In Fig.~\ref{superk}, we show the constraints on a light dark matter particle from Super-Kamiokande (adapted from Ref.~\cite{Hooper:2008cf}). From this figure, we conclude that if a dark matter particle responsible for the excess observed by CoGeNT annihilates primarily to neutrinos or tau leptons, it should have produced a flux of neutrinos observable to Super Kamiokande. If the dark matter annihilates significantly to other types of particles, however, it could produce the rates observed at CoGeNT and DAMA without coming into conflict with Super Kamiokande's observations. Alternatively, if the dark matter's annihilation cross section is highly suppressed (such as in the case of asymmetric dark matter, as discussed in Sec.~\ref{adm}), then these constraints do not apply.

Note that the results shown in Fig.~\ref{superk} differ somewhat from those found in Ref.~\cite{Hooper:2008cf}. This is because, in Ref.~\cite{Hooper:2008cf}, only scattering off of hydrogen and helium nuclei was included, and a simplified treatment of the velocity distribution was adopted~\cite{joakimpc}. The capture rates used to obtain the limits found in Fig.~\ref{superk} are in good agreement with the publicly available code DarkSUSY~\cite{darksusy}.

\subsection{Gamma Ray Searches From Dark Matter}

In generality, the flux of gamma rays from dark matter annihilations can be written as
\begin{equation}
\Phi_{\gamma}(E_{\gamma},\psi) = \frac{1}{2}\, \sigma |v| \, \frac{dN_{\gamma}}{dE_{\gamma}} \frac{1}{4\pi m^2_{\rm DM}} \int_{\rm{los}} \rho^2 dl,
\label{flux1}
\end{equation}
where, $\sigma |v|$ is the dark matter's annihilation cross section (times relative velocity), $\psi$ is the angle observed, $\rho$ is the dark matter density, and
the integral is performed over the line-of-sight being observed. $dN_{\gamma}/dE_{\gamma}$ is the gamma ray spectrum generated per WIMP annihilation, which depends on the mass of the dark matter particle and on its dominant annihilation channels. The integral in Eq.~\ref{flux1} depends on the distribution of dark matter, and is largest in the directions of the sky where very dense and relatively nearby concentrations of dark matter are present. The brightest source of gamma rays from dark matter annihilation is generally expected to be the region around the center of the Milky Way~\cite{gc}. Nearby dwarf spheroidal galaxies are also very promising sources for indirect detection~\cite{ds}.

To estimate the gamma ray flux from the Galactic Center region, we adopt a dark matter distribution which follows the commonly used Navarro-Frenk-White (NFW) profile~\cite{nfw}:
\begin{equation}
\rho(r) = \frac{\rho_0}{(r/R) [1 + (r/R)]^{2}} \,,
\label{profile}
\end{equation}
where $r$ is the distance from the Galactic Center, $R \sim 20$ kpc is the scale radius and $\rho_0$ is fixed by imposing that the dark matter density at the distance of the Sun from the Galactic Center is equal to the value approximately inferred by observations ($\sim\,$0.3 GeV/cm$^3$).

\begin{figure}[t]
\centering\leavevmode
\includegraphics[width=3.0in,angle=0]{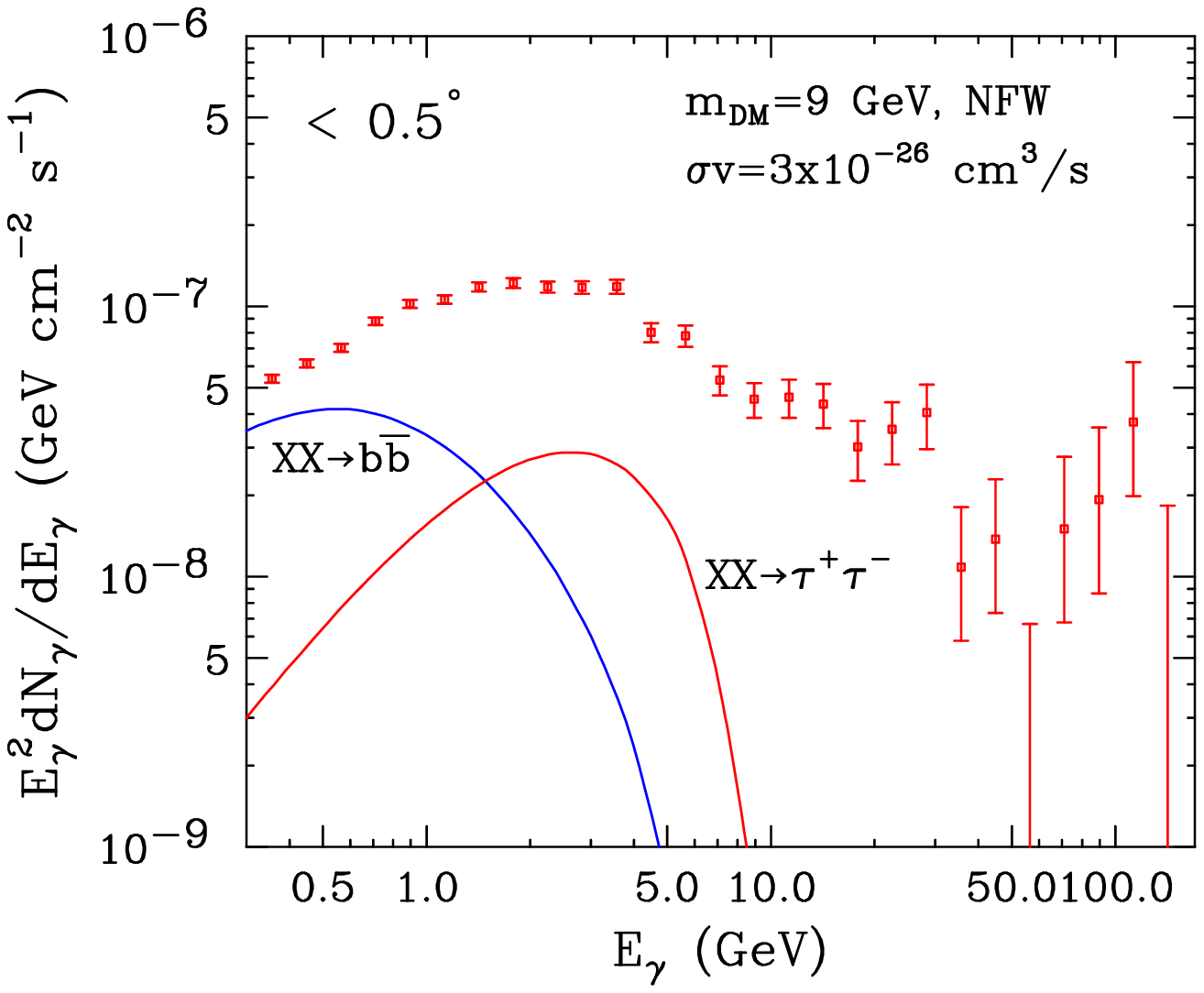}
\hspace{0.8cm}
\includegraphics[width=3.0in,angle=0]{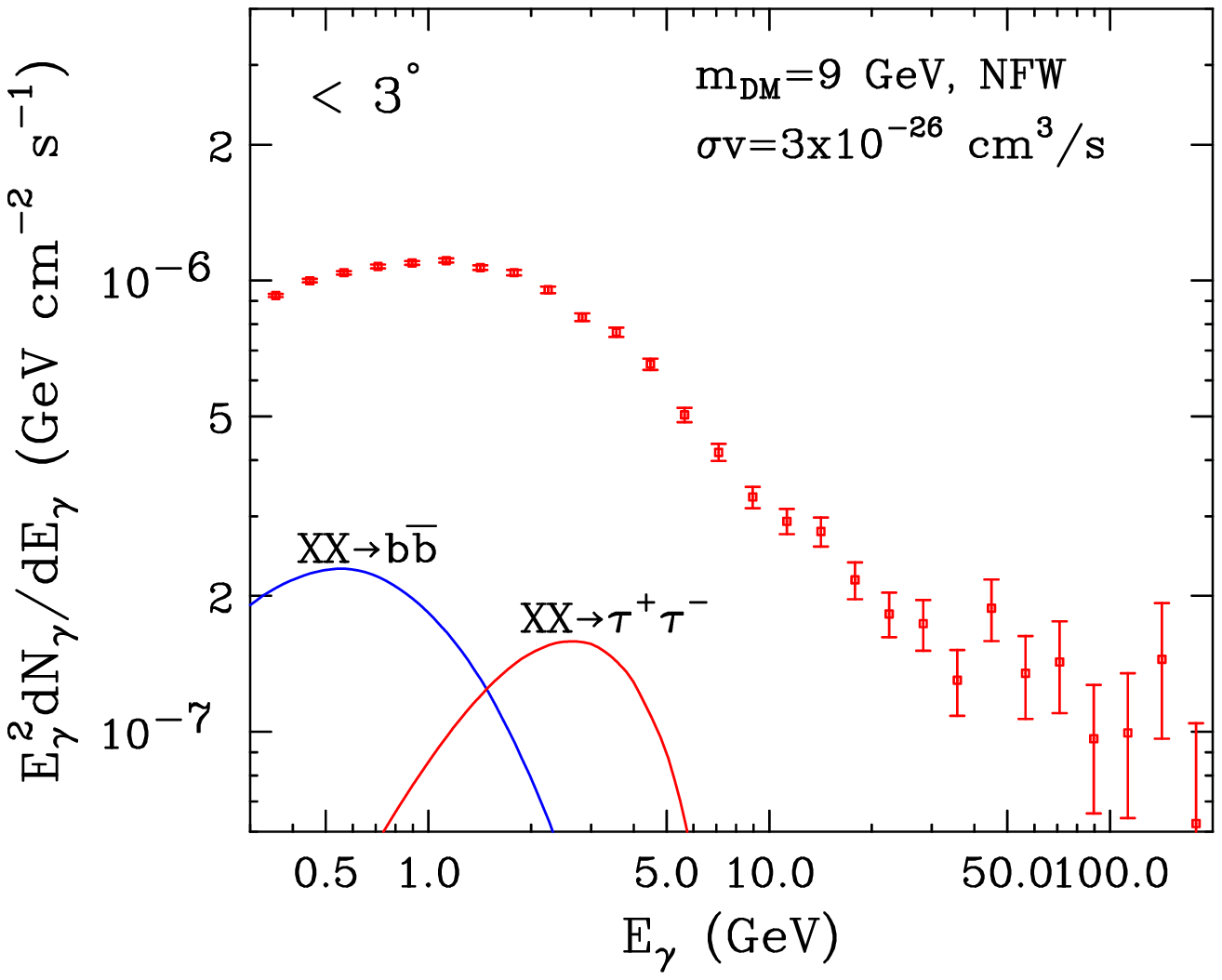}
\caption{The spectrum of gamma rays from 9 GeV dark matter particles annihilating in the inner Milky Way, compared to the measurements of the Fermi Gamma Ray Space Telescope (FGST). Here, we have assumed a dark matter distribution that follows an NFW profile, and an annihilation cross section of $\sigma v = 3\times 10^{-26}$ cm$^3$/s. See text for more details.}
\label{gamma}
\end{figure}

In Fig.~\ref{gamma}, we show the spectrum of gamma rays from dark matter annihilations in the inner Milky Way, and compare this to the spectrum measured by the Fermi Gamma Ray Space Telescope (FGST)~\cite{lisa}. The curves shown in each frame are for the case of a 9 GeV dark matter particle annihilating with a cross section of $\sigma v =  3 \times 10^{-26}$ cm$^3$/s (the value estimated for a typical thermal relic) to either $b\bar{b}$ or $\tau^+ \tau^-$. The left and right frames show the flux predicted from within the inner 0.5$^{\circ}$ and $3^{\circ}$ around the Galactic Center, respectively. The error bars shown correspond to the measurements of FGST over its first 14 months of observation (as shown in Ref.~\cite{lisa}). Each curve has been corrected to account for leakage due to FGST's finite point spread function (see Ref.~\cite{lisa} for details).

While the fluxes shown in Fig.~\ref{gamma} do not exceed those measured by FGST, they are quite close -- within a factor of 2-3 below 1-2 GeV. As more data is accumulated, and as the astrophysical backgrounds in this region become increasingly understood, FGST will likely become sensitive to 5-10 GeV dark matter particles with annihilation cross sections near that predicted for a simple thermal relic. The lone caveat to this conclusion is that, if the dark matter profile describing the inner galaxy does not contain a significant cusp (such as that described by the NFW model), then the gamma ray fluxes from dark matter annihilations could be considerably suppressed.

Recently, the FGST collaboration has reported its first limits on the gamma ray flux from a number of dwarf spheroidal galaxies, and used this information to constrain the properties of dark matter~\cite{fgstdwarf}. Although results are not presented for dark matter particles as light as 5-10 GeV (except in the special case of annihilations to $\mu^+ \mu^-$), the FGST collaboration find that dark matter particles with a 30 GeV mass and which annihilate to $b \bar{b}$ are restricted to $\sigma |v| \lsim 6\times 10^{-26}$ cm$^3$/s, assuming an NFW profile shape. Furthermore, the annihilation products of 5-10 GeV dark matter particles are predicted to contribute to the diffuse isotropic gamma ray flux at a level near the current sensitivity of FGST~\cite{Abazajian:2010sq}.

\subsection{Indirect Detection With Charged Cosmic Rays}

In contrast to most other astrophysical processes, dark matter annihilations are generally predicted to produce equal fluxes of matter and antimatter. An excess of antimatter in the cosmic ray spectrum relative to the predictions of astrophysical models could thus potentially constitute a signal of dark matter annihilations taking place in the halo of the Milky Way. 

Once injected into the interstellar medium, cosmic rays undergo a number of process which can alter their spectra as observed at Earth (for a review, see Ref.~\cite{SMPreview}). Such processes include deflection by and diffusion through the Galactic Magnetic Field, diffusive reacceleration, convection, and various energy loss processes resulting from interactions with gas, radiation fields, and magnetic fields. To confidently interpret the measurements of the cosmic ray spectrum, these processes must be understood and adequately modeled. 
   
At relatively high energies ($E \gsim 10$ GeV), many of these processes play only a relatively minor role. In particular, the effects of convection and reacceleration impact the cosmic ray spectrum much more strongly at low energies. This enables one to interpret the high energy measurements of PAMELA and (in the future) AMS-02 without introducing intractable uncertainties associated with the cosmic ray propagation model. For dark matter with a mass in the $\sim$10 GeV range, however, the resulting annihilation products will be of sufficiently low energy that they will likely be significantly effected by convection and reacceleration, as well as by interactions with the Solar System's magnetic field ({\it ie.}~solar modulation). Such effects may make it difficult to interpret these measurements within the context of indirect dark matter searches. That being said, lighter dark matter particles annihilate more often than heavier particles, leading to a greater flux of annihilation products to potentially detect~\cite{ferrer,Bottino:2007qg}. Without exploring these competing factors further, we simply comment that light dark matter particles are generally expected to produce large fluxes of antimatter in the $\sim$GeV region of the cosmic ray spectrum, but that the complicated nature of cosmic ray propagation over this range of energies may make it difficult to clearly identify such a contribution in existing or future measurements.

\subsection{Anomalous White Dwarf Heating}

Dark matter particles with large elastic scattering cross sections (such as those being considered here) can become captured by and annihilate in compact stars at very high rates. In relatively cool stars, such as very old white dwarfs, the energy injected from these annihilations could potentially compete with or exceed the star's luminosity, leading to observable effects~\cite{burners,burners2}. 

As a result of the very high densities of nuclei in white dwarfs, dark matter particles with elastic scattering cross sections of $\sigma\sim 10^{-40}$ cm$^2$ will scatter numerous times as they pass through the volume of the star. In this optically thick limit, dark matter particles will be gravitationally captured by a white dwarf at a rate given by
\begin{eqnarray}
\Gamma_c \approx \bigg(\frac{8 \pi}{3}\bigg)^{1/2} \frac{3 G \,R_{\rm WD}\, M_{\rm WD}\, \rho_{\rm DM}}{m_{\rm DM}\, \bar{v}},
\end{eqnarray}
where $R_{\rm WD}$ and $M_{\rm WD}$ are the radius and mass of the white dwarf, respectively, $\rho_{\rm DM}$ is the dark matter density in the region of the star, and $\bar{v}$ is the dark matter's velocity dispersion in that region. For a typical white dwarf ($R_{\rm WD}\approx 0.0083 R_{\odot}$, $M_{\rm WD} \approx 0.7 M_{\odot}$), this leads to a contribution to the star's luminosity given by~\cite{burners2}
\begin{eqnarray}
L&\approx& \Gamma_c \, m_{\rm DM} \nonumber \\
&\approx& 2.5 \times 10^{28} \, {\rm GeV/s} \, \bigg(\frac{\rho_{\rm DM}}{1\,{\rm GeV/cm}^3}\bigg) \bigg(\frac{220\,{\rm km/s}}{\bar{v}}\bigg) \nonumber \\
&\approx& 4 \times 10^{25} \, {\rm erg/s} \, \bigg(\frac{\rho_{\rm DM}}{1\,{\rm GeV/cm}^3}\bigg) \bigg(\frac{220\,{\rm km/s}}{\bar{v}}\bigg).
\label{lumdm}
\end{eqnarray}
For comparison, the Stefan-Boltzmann law predicts the luminosity of a blackbody to be
\begin{eqnarray}
L_{\rm BB} &=&4\pi R^2_{\rm WH} \sigma T^4_{\rm WD} \nonumber \\
&\approx& 1.9\times 10^{28} \,{\rm erg/s} \, \bigg(\frac{T_{\rm WD}}{3000 \, {\rm K}}\bigg)^4 \bigg(\frac{R_{\rm WD}}{0.0083 \, R_{\odot}}\bigg)^2.  
\label{lumwd}
\end{eqnarray}
Comparing Eqns.~\ref{lumdm} and~\ref{lumwd}, we see that a very old ($T\sim 3000$ K) white dwarf in a typical region of the galactic disk will have only on the order of 0.1\% of its luminosity generated by dark matter annihilations. In regions of high dark matter density and low dark matter velocity dispersion, however, dark matter annihilations can play a considerably greater role. The dark matter density and velocity dispersions within the inner $\sim$10-20 parsecs of dwarf spheroidal galaxies is typically on the order of $\rho_{\rm DM} \sim 10^2$ GeV/cm$^3$ and $\bar{v}\sim 10$ km/s, respectively~\cite{Walker:2007ju}. In such an environment, dark matter of the type required to produce the excesses observed by CoGeNT and DAMA will prevent white dwarf stars from cooling below $\sim 5000$ K~\cite{burners2}. This provides an opportunity for future telescopes not only to confirm a dark matter interpretation of the CoGeNT result, but also to potentially map the density and velocity dispersion of dark matter in regions such as dwarf spheroidal galaxies.

\section{Conclusions}

In this paper, we have discussed the recent excess of low energy events reported by the CoGeNT collaboration, and considered the possibility that these events, as well as the annual modulation observed by DAMA, are the result of a light ($m_{\rm DM}\sim 5-10$ GeV) dark matter particle. We find that a common dark matter interpretation may be compatible with the CoGeNT and DAMA signals, as well as the modest excess recently reported by CDMS. We find that this interpretation is also consistent with the null results of XENON10 and the CDMS silicon analysis.

Dark matter with the properties (mass and elastic scattering cross section) required to produce the CoGeNT and DAMA excesses can appear within a variety of theoretical frameworks. From a model independent perspective, we note that a $\sim$5-10 GeV scalar dark matter particle with approximately universal scalar couplings to Standard Model fermions will naturally produce approximately the observed event rates if its couplings are fixed to obtain the desired thermal relic abundance of dark matter. Fermionic dark matter particles are also a viable possibility, although multiple types of interactions appear necessary if such a particle is also to satisfy relic density constraints. Within the context of specific theoretical models, we have discussed both Asymmetric Dark Matter models and neutralinos within extended supersymmetric models, such as the NMSSM. Asymmetric Dark Matter is especially attractive as an explanation for the CoGeNT and DAMA signals, as in such scenarios the dark matter's mass is related to the proton's mass and is generally predicted to be roughly in the range of $m_{\rm DM}\sim$1-10 GeV. 

We have also discussed the implications of the CoGeNT and DAMA excesses for other dark matter search strategies, and find that the prospects for indirect detection are very favorable in many such scenarios. In particular, existing limits from low-threshold neutrino detectors, such Super Kamiokande, already constrain the dominant annihilation channels of such a WIMP. Searches for dark matter annihilations products with the Fermi Gamma Ray Space Telescope (FGST) currently have a level of sensitivity that is very close to being able to test dark matter in a form of a $\sim$5-10 GeV thermal relic.

\section*{Acknowledgements} We would like to thank Juan Collar for his assistance in understanding the details of the recent CoGeNT result. We also thank B. 
Feldstein, A. Pierce, and P. Sorensen for helpful discussions. DH is supported by the US Department of Energy, including grant DE-FG02-95ER40896, and by NASA grant NAG5-10842. 
ALF and KMZ thank the KITP for their hospitality at the workshop 
``Direct, Indirect and Collider Signals of Dark Matter,'' where parts of this work were initiated.  ALF is supported by 
DOE grant DE-FG02-01ER-40676 and NSF CAREER grant PHY-0645456.

\end{document}